\documentclass[preprint,12pt]{elsart}

\usepackage{graphicx}

\usepackage{amssymb}
\usepackage{amsmath}
\usepackage{dsfont}
\usepackage{amsfonts}
\usepackage{cancel}
\usepackage{color}
\usepackage{epstopdf}

\journal{Nuclear Physics A}

\begin{document}

\begin{frontmatter}



\title{Calculations of antiproton-nucleus quasi-bound states using the Paris $\bar{N}N$ potential}


\author{Jaroslava Hrt\'{a}nkov\'{a} and Ji\v{r}\'{\i} Mare\v{s}}

\address{Nuclear Physics Institute, 250 68 \v{R}e\v{z}, Czech Republic}

\begin{abstract}
An optical potential constructed using the ${\bar p}N$ scattering amplitudes derived from the 
2009 version of the Paris ${\bar N}N$ potential is applied in  calculations of 
${\bar p}$ quasi-bound states in selected nuclei across the periodic table. A proper self-consistent procedure 
for treating energy dependence of the amplitudes in a nucleus appears crucial for evaluating 
${\bar p}$ binding energies and widths. Particular attention is paid to the role of $P$-wave amplitudes. 
While the $P$-wave potential nearly does not affect calculated ${\bar p}$ binding energies, it 
reduces considerably the corresponding widths. 
The Paris $S$-wave potential supplemented 
by a phenomenological $P$-wave term yields in dynamical calculations ${\bar p}$ binding energies 
$B_{\bar p}\approx 200$~MeV  
and widths $\Gamma_{\bar p}\sim~200 - 230$~MeV, which is very close to the values obtained within the 
RMF model consistent with ${\bar p}$-atom data.    
\end{abstract}


\end{frontmatter}

\section{Introduction}
\label{intro}
The elastic part of an ${\bar N}N$ potential constructed from a boson exchange $NN$ potential using the G-parity 
transformation is strongly attractive~(see, e.g.~\cite{machlaidt88}). This fact stimulated  
speculations about existence of ${\bar p}$-nuclear bound states~\cite{Wong,Baltz,burvenich}. 
The ${\bar p}N$ and ${\bar p}$-nucleus interactions, as well as their capability of forming corresponding 
bound states have been explored extensively in LEAR experiments at CERN~\cite{antiNN interaction, KlemptAnnih}.
   Complementary information about the ${\bar p}$ optical potential 
near threshold was acquired within the study of strong interaction energy shifts and widths in antiprotonic 
atoms~\cite{batty,friedman,mares}.   
Analyses of ${\bar p}$-atom data and  ${\bar p}$ scattering off nuclei at low energies 
disclosed that the antiproton interaction with a nucleus is dominated by ${\bar p}$ 
annihilation which governs propagation of the antiproton in nuclear matter. 
It was found that the experimental data could be well fitted by a ${\bar p}$-nuclear optical 
potential, imaginary part of which greatly outweighs the strongly attractive real part. 
However, if the antiproton is deeply bound in the nuclear medium the situation might change.  
In fact, the phase space for ${\bar p}$ annihilation products gets considerably reduced, which 
could lead to relatively long ${\bar p}$ lifetime in nuclear matter~\cite{Mishustin}. 
Nonetheless, no definite evidence of forming a ${\bar p}N$ or ${\bar p}$-nucleus quasi-bound state 
has been reported so far.

The study of ${\bar p}$ interactions with a nucleon and the nuclear medium is still topical. 
One example worth mentioning is the very recent analysis of $J/\Psi$ events collected by 
the BESIII experiment which supports the existence of either a ${\bar p}p$ molecule-like state or a bound 
state~\cite{BESIII}. It is to be noted that one of the observed resonant states, $X(1835)$,  
was described by the 2009 version of the Paris ${\bar N}N$ potential, assuming that it originates from 
a ${\bar p}p$ bound state~\cite{paris2,BESIII}. Furthermore, 
the knowledge of ${\bar p}$-nucleus interaction at various densities and under different 
kinematical conditions will be utilized and further expanded in forthcoming experiments with ${\bar p}$-beams 
at FAIR~\cite{FAIR}. Simulations of the considered processes in a wide range of ${\bar p}$-beam momenta,    
providing experimentalists with valuable hints, are being performed within the Giessen Boltzmann-Ueling-Uhlenbeck 
(GiBUU) transport model~\cite{gibuu} in which the ${\bar p}$ potential in nuclear matter serves as input. 

Properties of ${\bar p}$-nuclear quasi-bound states have been calculated within the Relativistic 
Mean Field (RMF) model~\cite{burvenich,Mishustin,larionovPRC08,lmsgPRC10,lenske,hmNPA16} using the 
G-parity transformation of coupling constants involved. 
A scaling factor representing departure from G-parity together with a phenomenological imaginary part 
were introduced to construct an optical potential consistent with experimental data.  
In Ref.~\cite{hmNPA16}, the $\bar p$ annihilation was treated dynamically and fully self-consistently, 
taking into account the reduced phase-space for annihilation products as well as compression 
of the nuclear core caused by the antiproton. 
Though the calculated ${\bar p}$ widths in the nuclear medium were found to be suppressed significantly, 
they remained considerable. 
Recently, Gaitanos et al. have developed a non-linear derivative (NLD) model~\cite{NLD1,NLD3}.  
It incorporates momentum dependence of mean 
fields acting on ${\bar p}$ which yields the depth of the ${\bar p}$-nucleus potential 
in accord with experimental data, without introducing any additional scaling factor.   
 
Several microscopic ${\bar N}N$ potential models, such as those based on one- and two-pion exchange~\cite{paris2,ZT} 
or chiral EFT~\cite{KHM14,H17} have been developed recently. 
Friedman et al.~\cite{friedmanNPA15} confronted the 2009 version of the Paris 
${\bar N}N$ potential~\cite{paris2} with the ${\bar p}$-atom data across the periodic table and 
antinucleon interactions with nuclei up to 400~MeV/c, including elastic scattering and annihilation 
cross sections. Their analysis revealed necessity to include the $P$-wave part of the ${\bar p}N$ 
interaction to make the real ${\bar p}$ potential attractive in the relevant low density region 
of the nucleus, as required by experiment. However, it was found that the Paris $S$-wave potential 
supplemented by the contribution of the Paris $P$-wave amplitudes fails to achieve reasonable 
fit to  ${\bar p}$ atom data. On the other hand, the Paris $S$-wave potential with a purely 
phenomenological $P$-wave term accounts well for the data on the low-density, near-threshold 
${\bar p}$-nucleus interaction. From this point of view, it is tempting to apply the $S$-wave 
amplitudes derived from the Paris ${\bar N}N$ potential and either the Paris or phenomenological 
$P$-wave amplitudes to the description of ${\bar p}$ interactions in the nuclear interior, i.e.,  
farther down below threshold and at higher nuclear densities.
    
In the present work we employ the 2009 version of the Paris ${\bar N}N$ model in the construction 
of an optical potential which is then used in calculations of ${\bar p}$-nuclear quasi-bound states for 
the first time. 
We demonstrate the role of a proper self-consistent treatment of the energy dependence of 
scattering amplitudes involved. The adopted procedure for evaluating the sub-threshold energy shift 
has been applied before in calculations of kaonic and ${\bar p}$ atoms, as well as $K^-$-, $\eta$- and 
${\bar p}$-nuclear states~\cite{hmNPA16,friedmanNPA15,s,kaony,kaonic atoms1,etaS1,etaS2,fg17,hmPLB17,hmPRC17}.    
We take into account the $P$-wave part of the ${\bar p}N$ potential aiming at 
exploring its impact on calculated ${\bar p}$ binding energies and widths. 
Finally, we compare present results with those obtained within the RMF approach constrained by   
${\bar p}$-atom data~\cite{hmNPA16}. 
 
The paper is organized as follows. In Section 2, we briefly describe applied methodology. 
We present construction of the in-medium ${\bar p}N$ $S$-wave amplitudes from the free-space 
amplitudes derived within the Paris $\bar{N}N$ potential.    
We discuss a self-consistent procedure for treating 
the energy dependence of the amplitudes and construct a  relevant ${\bar p}$ optical potential. 
Moreover, we introduce the $P$-wave interaction term which supplements the $S$-wave part of 
the potential.  
In Section~3, we present selected results of our calculations of ${\bar p}$ quasi-bound 
states in various nuclei across the periodic table, illustrating dynamical effects 
in the nuclear core caused by the antiproton and the role of the $P$-wave part of the ${\bar p}N$ potential.   
 Summary of the present study is given in Section~4.  

\section{Model}
The binding energy $B_{\bar{p}}$ and width $\Gamma_{\bar{p}}$ of a $\bar{p}$ quasi-bound state in a nucleus are obtained 
by solving the Dirac equation
\begin{equation} \label{DiracEq}
[-i\vec{\alpha} \cdot \vec{\nabla} +\beta m_{\bar{p}}+V_{\text{opt}}(r)]\psi_{\bar{p}}=\epsilon_{\bar{p}} \psi_{\bar{p}},
\end{equation}
where $m_{\bar{p}}$ is the mass of the antiproton,  $\epsilon_{\bar{p}}=-B_{\bar{p}} - {\rm i}\Gamma_{\bar{p}}/{2}$, $(B_{\bar{p}} > 0)$, and $V_{\text{opt}}(r)$ is a complex optical potential which enters 
the Dirac equation as a time component of a 4-vector~\footnote{As a test, we solved the Schr\"{o}dinger equation 
for the same potential $V_{\text{opt}}(r)$ and got $\bar p$ energies and widths which differ by less than 1~MeV from 
those obtained by solving Eq.~\eqref{DiracEq}.}. 

\subsection{S-wave interaction}
First, we consider only the $S$-wave optical potential which is constructed in a '$t\rho$' form as follows: 
\begin{equation} \label{SoptPot}
 2E_{\bar{p}}V^{\rm S}_{\text{opt}}(r)=-4\pi \left(F_0\frac{1}{2}\rho_p(r) 
+ F_1\left(\frac{1}{2}\rho_p(r)+\rho_n(r) \right)\right)~,
\end{equation}
where $E_{\bar{p}}=m_{\bar{p}}-B_{\bar{p}}$, $F_0$ and $F_1$ are isospin 0 and 1 in-medium amplitudes in 
the ${\bar p}$-nucleus frame, and 
$\rho_p(r)$ [$\rho_n(r)$] is the proton (neutron) density distribution calculated 
within the RMF model NL-SH \cite{nlsh}~\footnote
{The NL-SH model contains non-linear scalar self-interactions comprising of the cubic and quadratic terms in the 
$\sigma$ field. The model has proven successful in reproducing the binding energies and charge radii of nuclei, as well 
as neutron-skin thickness. In addition, it describes well saturation properties of nuclear matter, such as the 
binding energy per nucleon $a_{\rm v} = -16.33$~MeV, nuclear matter density $\rho_0 = 0.146$~fm$^{-3}$, and 
compressibility $K = 355$~MeV. 
}. The amplitudes $F_0$ and $F_1$ entering Eq.~\eqref{SoptPot} are constructed from the free-space $\bar{p}N$ amplitudes in the two-body frame 
using the multiple scattering approach of Wass et al. \cite{wrw} (WRW) which accounts for Pauli correlations in the nuclear medium
\begin{equation}\label{wrw}
F_{1}=\frac{\frac{\sqrt{s}}{m_N}f^{\rm S}_{\bar{p}n}(\delta \sqrt{s})}{1\!+\!\frac{1}{4}\xi_k \frac{\sqrt{s}}{m_N} f^{\rm S}_{\bar{p}n}
(\delta \sqrt{s}) \rho (r)}~,\;\; F_{0}=\frac{\frac{\sqrt{s}}{m_N}[2f^{\rm S}_{\bar{p}p}(\delta \sqrt{s})\! -\! f^{\rm S}_{\bar{p}n}
(\delta \sqrt{s})]}{1\!+\!\frac{1}{4}\xi_k \frac{\sqrt{s}}{m_N}[2f^{\rm S}_{\bar{p}p}(\delta \sqrt{s})\!  
-\! f^{\rm S}_{\bar{p}n}(\delta \sqrt{s})] \rho (r)}~. 
\end{equation} 
Here, $f^{\rm S}_{\bar{p}n}$ ($f^{S}_{\bar{p}p}$) denotes the free-space $\bar{p}n$ (${\bar{p}p}$) $S$-wave 
two-body cm scattering amplitude 
as a function of 
$\delta \sqrt{s}=\sqrt{s}-E_{\text{th}}$, where $s$ is the Mandelstam variable and $E_{\text{th}}=m_N+m_{\bar{p}}$.   
The factor 
$\sqrt{s}/m_N$ transforms the amplitudes from the two-body frame to the $\bar{p}$-nucleus frame. 
The nuclear density distribution $\rho(r) = \rho_p(r)+\rho_n(r)$ and the Pauli correlation 
factor $\xi_k$ is defined as follows
\begin{equation} \label{xi}
 \xi_k=\frac{9\pi}{k_{\rm F}^2}\,\left(4\! \int_0^{\infty} \frac{dr}{r} \exp(ikr)j_1^2(k_{\rm F} r)\right)\; ,
\end{equation}
where $j_1(k_{\rm F} r)$ is the spherical Bessel function, $k_{\rm F}$ is the Fermi momentum, 
and $k=\sqrt{(\epsilon_{\bar{p}}+m_{\bar{p}})^2-m_{\bar{p}}^2}$. 
The integral in Eq.\eqref{xi} can be solved analytically. The resulting expression is of the form
\begin{equation}
 \xi_k=\frac{9\pi}{k_{\rm F}^2} \left[1- \frac{q^2}{6}+\frac{q^2}{4}\left(2+ \frac{q^2}{6}\right)
\ln\left(1+\frac{4}{q^2}\right)- \frac{4}{3}q \left( \frac{\pi}{2}- \arctan \left( \frac{q}{2} \right) \right) \right]~,
\end{equation}
where $q=-ik/k_{\rm F}$.

The free-space $S$-wave scattering amplitudes are derived from the 2009 version of the 
Paris $\bar{N}N$ potential \cite{paris2}. The $\bar{p}n$ and $\bar{p}p$ amplitudes are expressed as appropriate 
mixtures of isospin $T=0$ and $T=1$ $\bar{N}N$ amplitudes, evaluated as angular momentum averages of fixed-$T$ 
amplitudes~\cite{friedmanNPA15}. 

\begin{figure}[t]
\begin{center} 
\includegraphics[width=0.47\textwidth]{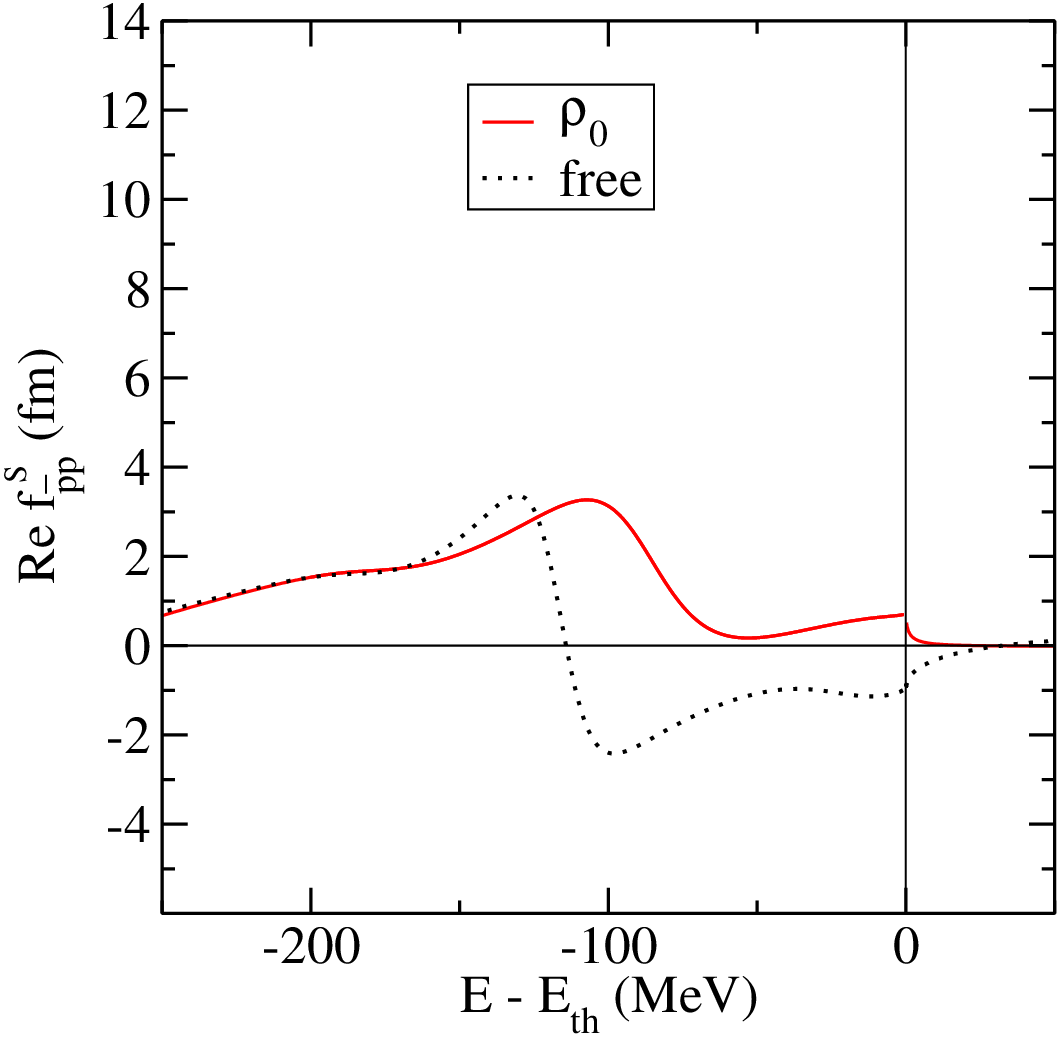} \hspace{10pt}
\includegraphics[width=0.47\textwidth]{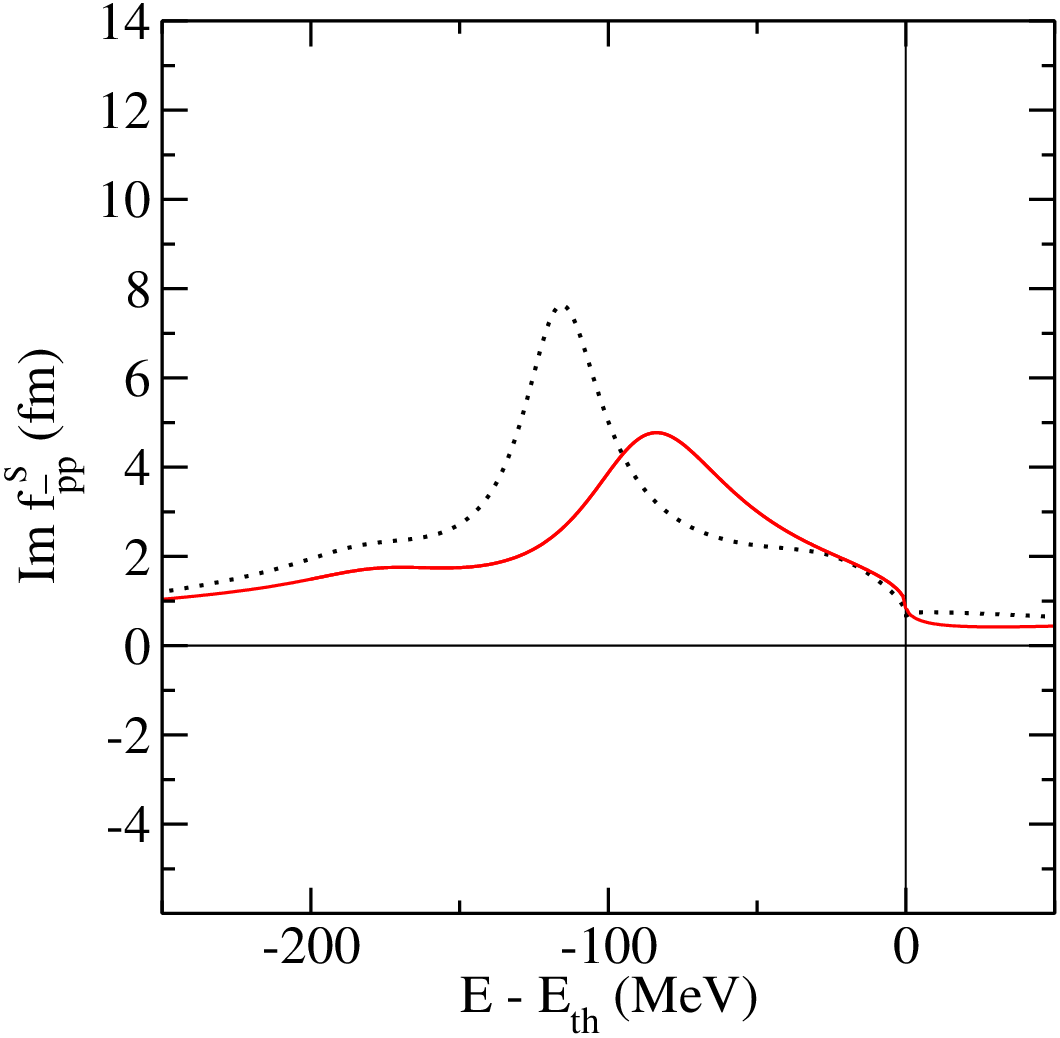} \\[10pt]
\includegraphics[width=0.47\textwidth]{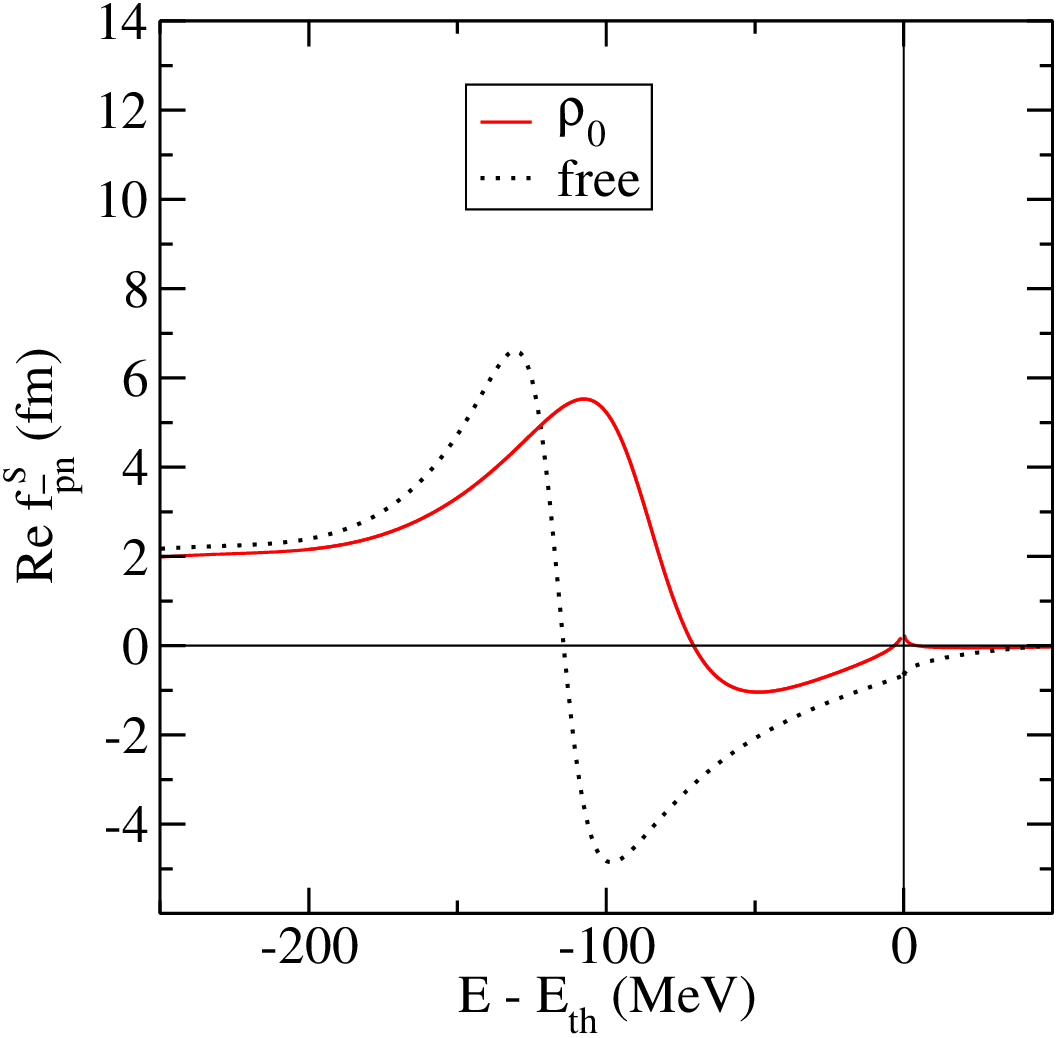} \hspace{10pt}
\includegraphics[width=0.47\textwidth]{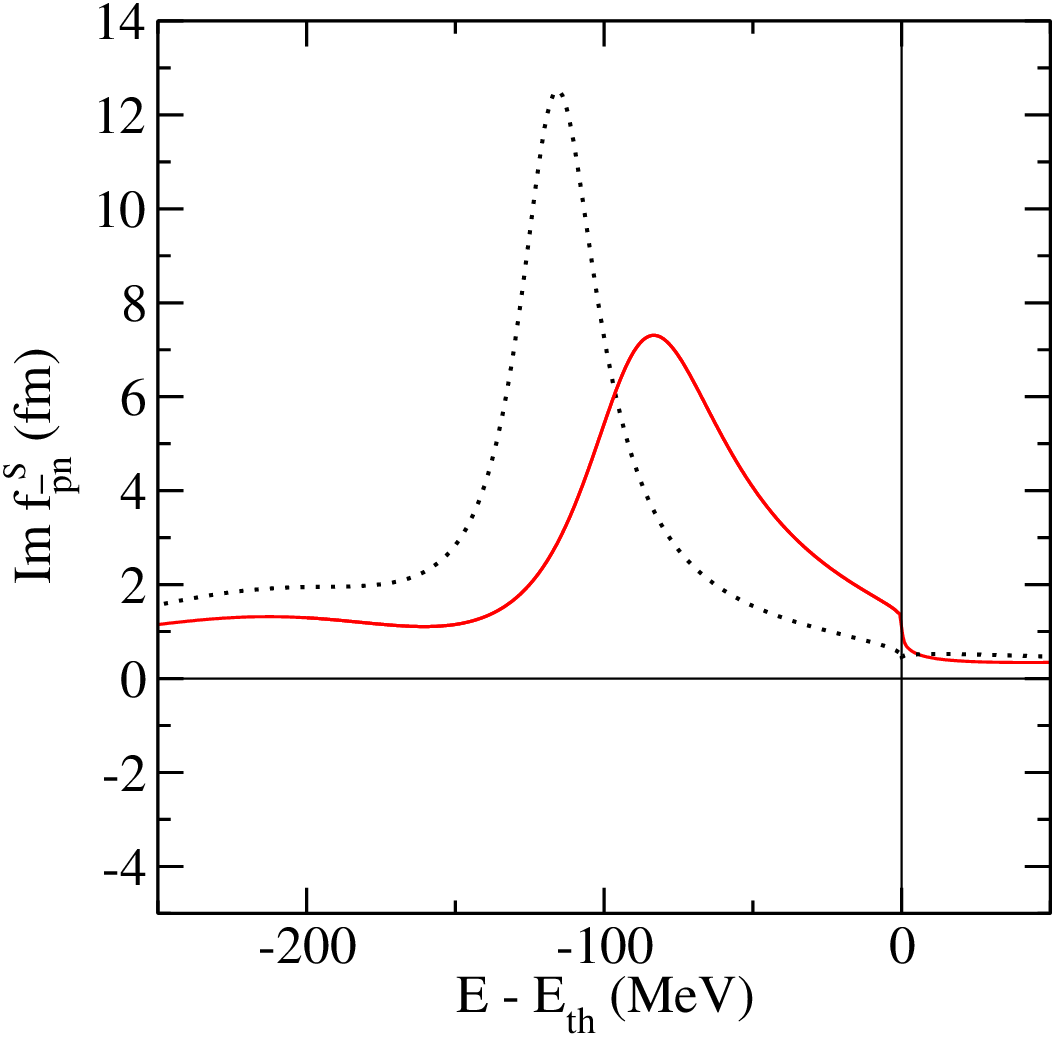}
\caption{\label{fig.:1}Energy dependence of the Paris 09 $\bar{p}p$ (top) and $\bar{p}n$ (bottom) 
$S$-wave two-body cm amplitudes: in-medium (Pauli blocked) amplitudes for $\rho_0=0.17$~fm$^{-3}$ (solid line) 
are compared with the free-space amplitude (dotted line).}
\end{center}
\end{figure}
In Fig.~\ref{fig.:1} the free-space $\bar{p}p$ (top panel) and $\bar{p}n$ (bottom panel) amplitudes  
plotted as a function of energy are compared with the in-medium amplitudes at saturation density 
$\rho_0=0.17$~fm$^{-3}$. Both free and WRW modified amplitudes manifest strong energy dependence for 
$\delta\sqrt{s} = E-E_{\rm th}\leq 0$. 
While the in-medium $\bar{p}p$ amplitude is attractive in the entire energy range 
below threshold,  the real part of the in-medium $\bar{p}n$ amplitude is attractive for $\delta\sqrt{s} \leq -70$~MeV 
 and with slightly repulsive dip near threshold. 
The peaks of both in-medium amplitudes are 
lower in comparison with the free-space amplitudes and shifted by $\approx 30$~MeV towards the $\bar{p}N$ threshold. 

We explored the effect of the WRW procedure on $\bar{p}$ binding energies and widths and performed, out of curiosity,  
also  calculation with the free-space $S$-wave amplitudes. 
In Table~\ref{Tab.:WRW} we present $1s$ $\bar{p}$ binding energies and widths in $^{208}$Pb calculated with the free-space 
amplitudes and WRW modified (in-medium) amplitudes, using static RMF densities. 
The in-medium modifications significantly reduce the $\bar{p}$ widths whereas the $\bar{p}$ binding energies are 
affected only slightly. This could be anticipated upon closer inspection of Fig.~\ref{fig.:1}. The differences 
between the free-space and WRW-modified real ${\bar p}N$ amplitudes at $\delta \sqrt{s} = E-E_{\rm th} \sim -200$~MeV 
(which is the energy shift relevant to static calculations) is almost negligible (see left panels). 
On the other hand, the imaginary amplitudes (right panels) are evidently reduced at $\delta \sqrt{s} \sim -200$~MeV 
when in-medium modifications are taken into account. Consequently, the ${\bar p}$ widths are reduced as well. 

\begin{table}[t]
\caption{$1s$ $\bar{p}$ binding energies and widths (in MeV) in $^{208}$Pb, calculated using static RMF densities with 
the free-space (free) and in-medium (WRW) $S$-wave amplitudes.} 
\vspace*{10pt}
\begin{center}
 \begin{tabular}{l|cc}
\hline
  & free & WRW \\ \hline
 $B_{\bar{p}}$ (MeV) & 184.8  & 188.6\\
 $\Gamma_{\bar{p}}$ (MeV) & 318.5 & 233.8\\
\hline 
 \end{tabular}
\vspace*{10pt}
\end{center}
\label{Tab.:WRW}
\end{table}

The energy argument of the ${\bar{p}N}$ scattering amplitudes is expressed by Mandelstam variable
\begin{equation} \label{s}
s=(E_N + E_{\bar{p}})^2 - (\vec{p}_N + \vec{p}_{\bar{p}})^2~,
\end{equation} 
where $E_N=m_N - B_{N\rm av}$ with $B_{N\rm av}=8.5$~MeV being the average 
binding energy per nucleon. In the two-body c.m. frame $\vec{p}_N + \vec{p}_{\bar{p}} = 0$ and Eq.~\eqref{s} 
reduces to
\begin{equation} \label{Eq.:M}
\sqrt{s}_{\rm M}=~m_{\bar{p}}+m_{N}-B_{\bar{p}}-B_{N\rm av}~.
\end{equation}
However, when the interaction of the antiproton with a nucleon takes place in a nucleus, the momentum 
dependent term in Eq.~\eqref{s} does not vanish and gives rise to an additional downward energy shift~\cite{s}.   
Taking into account averaging over the angles $(\vec{p}_N + \vec{p}_{\bar{p}})^2 \approx \vec{p}_N^{~2}+\vec{p}_{\bar{p}}^{~2}$,
Eq.~\eqref{s} can be rewritten as 
\begin{equation} \label{Eq.:J}
 \sqrt{s}_{\rm J}= E_{\rm th} \left(\!1-\frac{2(B_{\bar{p}} + B_{N\rm av})}{E_{\rm th}} + \frac{(B_{\bar{p}}+ B_{N\rm av})^2}{E_{\rm th}^2} 
- \frac{T_{\bar{p}}}{E_{\rm th}} - \frac{T_{N\rm av}}{E_{\rm th}}\!\right)^{1/2}~,
\end{equation}
where $T_{N\rm av}$ is the average kinetic energy per nucleon and $T_{\bar{p}}$ represents the kinetic energy of the antiproton. 
The kinetic energies were calculated as the corresponding expectation values of the kinetic energy operator 
$\hat{T}=-\frac{\hbar^2}{2 m_N} \triangle$.

We note that the ${\bar p}$ binding energy $B_{\bar p}$ appears as an argument in the expression for 
$\sqrt{s}$, which in turn serves as an argument for $V_{\rm opt}$ 
in Eq~\eqref{DiracEq}. 
Therefore, $\sqrt{s}$ has to be determined self-consistently, namely its 
value obtained by solving the Dirac 
equation~\eqref{DiracEq} should agree with the value of $\sqrt{s}$ which serves as input in 
Eqs.~\eqref{SoptPot} and \eqref{wrw}.
An additional self-consistency scheme has to be considered in dynamical calculations: The RMF densities  
entering the expression~\eqref{SoptPot} for $V_{\rm opt}$ are modified by the ${\bar p}$ bound in a 
nucleus and thus by the solution of the Dirac equation~\eqref{DiracEq}.   

\begin{figure}[t]
\begin{center} 
\includegraphics[width=0.47\textwidth]{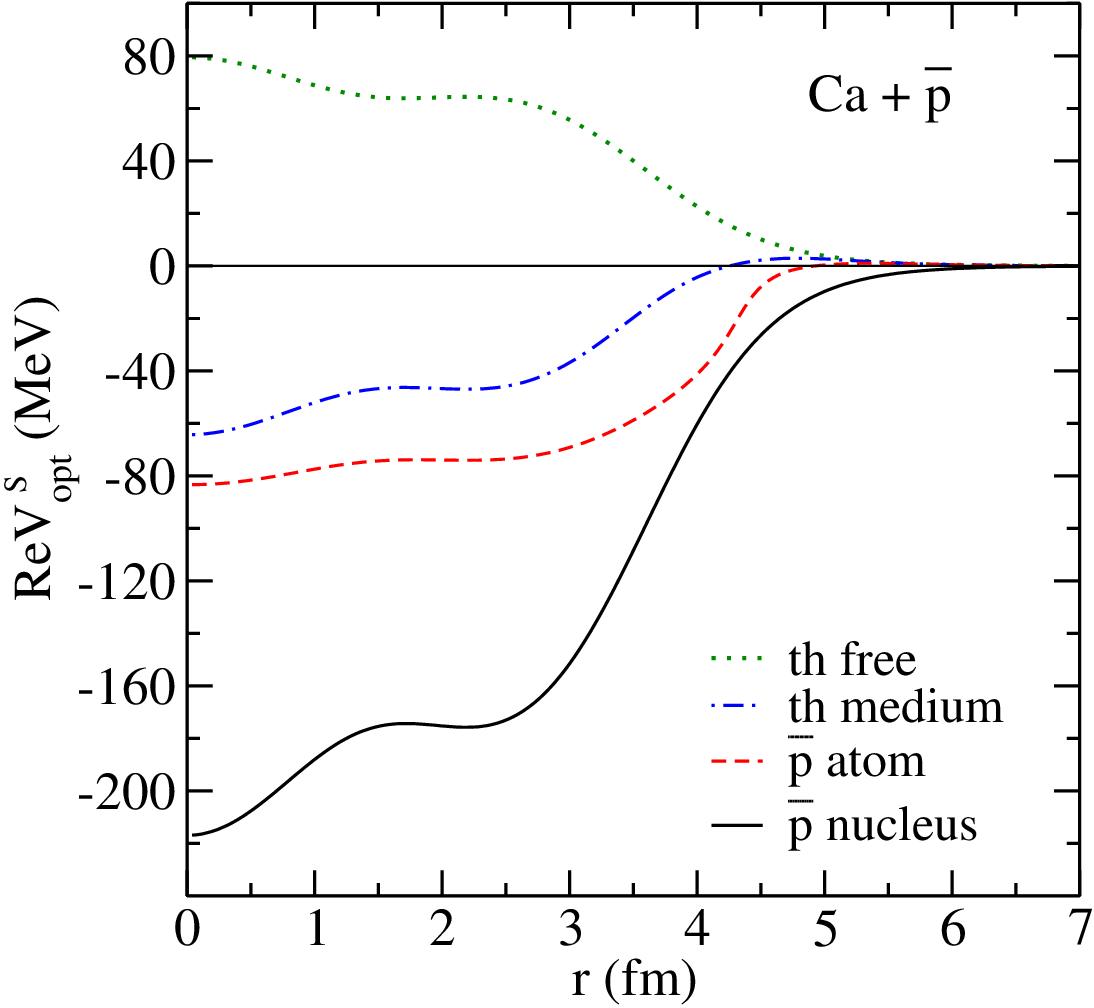} \hspace{10pt}
\includegraphics[width=0.47\textwidth]{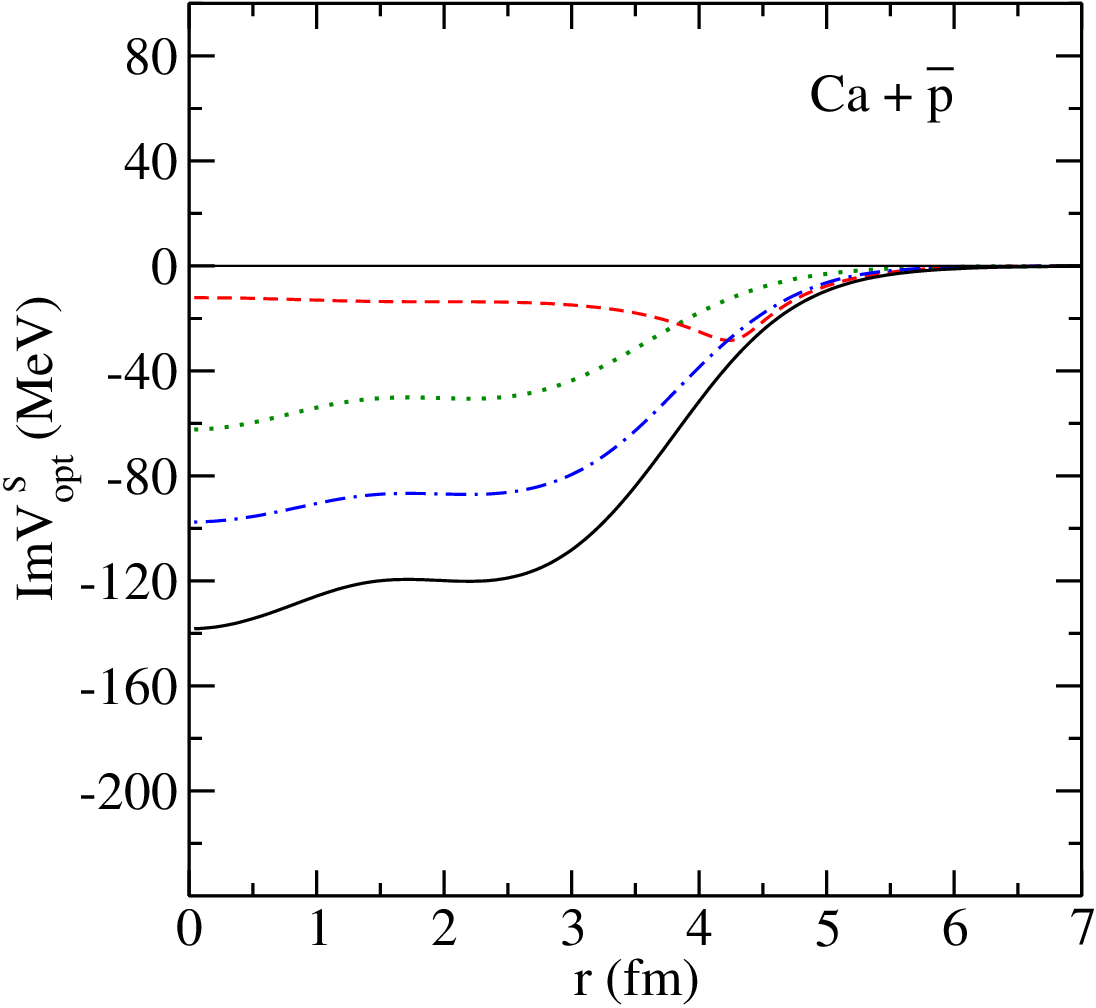}
\caption{\label{fig.:2}The potential felt by $\bar{p}$ at threshold (`th medium'), in the $\bar{p}$ atom and $\bar{p}$ nucleus, 
calculated for $^{40}$Ca+$\bar{p}$ with in-medium Paris $S$-wave amplitudes and static RMF densities. 
The ${\bar p}$ potential calculated using free-space amplitudes at threshold is shown for comparison (`th free').}
\end{center}
\end{figure}

The $\bar{p}N$ amplitudes are strongly energy and density dependent, as was shown in Fig.~\ref{fig.:1}. Consequently, 
the depth and shape of the $\bar{p}$-nuclear potential depend greatly on the energies and densities 
pertinent to the processes 
under consideration. This is demonstrated in Fig.~\ref{fig.:2} where we 
present the $\bar{p}$ potential in $^{40}$Ca calculated using the free-space amplitudes at threshold and 
in-medium Paris $S$-wave amplitudes in three different energy 
regions: At threshold, for $\delta\sqrt{s}=0$, 
the $\bar{p}$ potential constructed using the free space amplitudes (denoted by `th free') has a repulsive real 
part and fairly absorptive imaginary part. When medium modifications of the amplitudes are taken into account (`th medium'), the $\bar{p}$ potential becomes attractive and more absorptive.  
At energies relevant to ${\bar p}$ atoms the $\bar{p}$ potential, constructed following Ref.~\cite{friedmanNPA15}, is more attractive and weakly absorptive.  
At energies relevant to  $\bar{p}$ nuclei ($\sqrt{s}_{\rm J}$), the $\bar{p}$ potential is strongly attractive, however, 
also strongly absorptive.  Clearly, proper self-consistent evaluation of the energy shift $\delta \sqrt{s}$ is 
crucial. 

\subsection{P-wave interaction}

\begin{figure}[t]
\begin{center}
\includegraphics[width=0.47\textwidth]{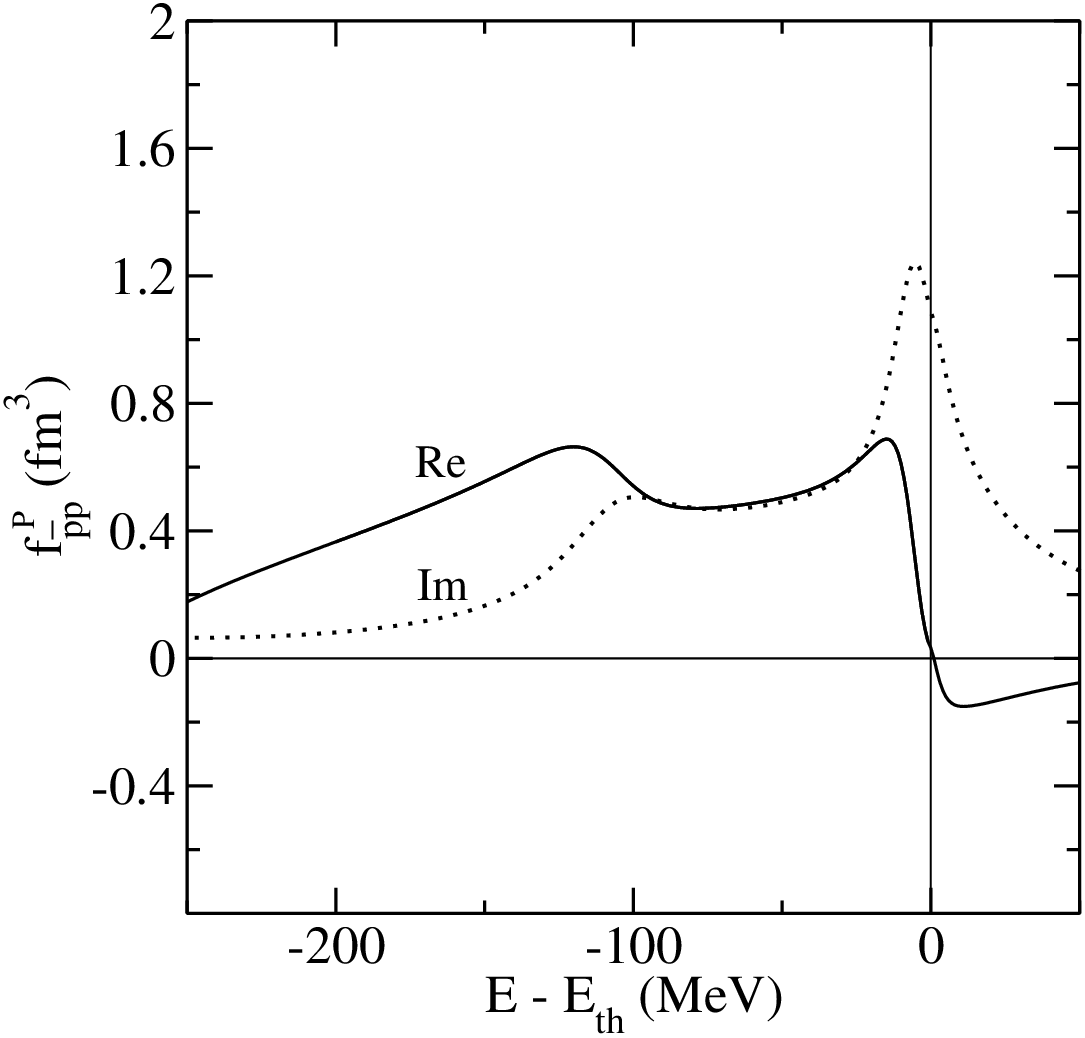} \hspace{10pt}
\includegraphics[width=0.47\textwidth]{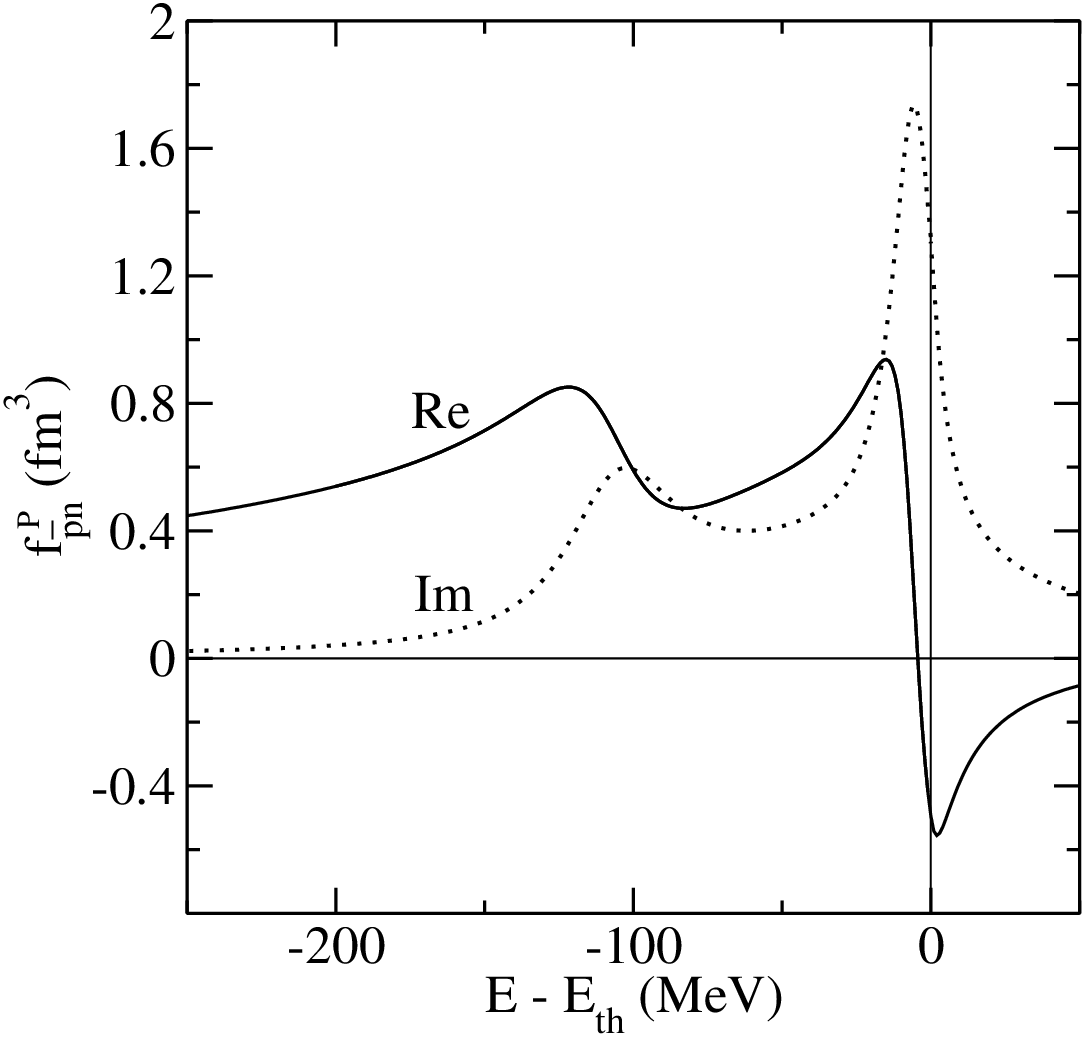}
\caption{\label{fig.:3}Energy dependence of the Paris 09 $\bar{p}p$ (left) and $\bar{p}n$ (right) 
$P$-wave free-space amplitudes.}
\end{center}
\end{figure}

Recent calculations of ${\bar p}$ atoms and scattering of 48~MeV antiprotons~\cite{friedmanNPA15}  
showed that a sizable contribution from the $P$-wave part of the ${\bar p}N$ interaction is needed 
to get reasonable description of the experimental data.  
In order to examine the effect of the $P$-wave interaction on the binding energies and widths of ${\bar p}$-nuclear 
quasi-bound states, we supplement the $S$-wave optical potential in  Eq.~\eqref{SoptPot} 
[$q(r) = 2E_{\bar{p}}V^{\rm S}_{\text{opt}}(r)$] by a gradient 
term which stands for the $P$-wave interaction~\cite{friedmanNPA15, satchlerAP96, ericsonAP66}
\begin{equation}\label{SPpot}
 2E_{\bar{p}}V_{\text{opt}}(r)=q(r)+3 \vec{\nabla} \cdot \alpha(r) \vec{\nabla}\; .
\end{equation}
The factor $2l+1 = 3$ in the $P$-wave part is introduced to match the normalization of the Paris $\bar{N}N$ 
scattering amplitudes and 
\begin{equation}
 \alpha(r) = 4 \pi \frac{m_N}{\sqrt{s}}\left(f^{P}_{\bar{p}p}(\delta \sqrt{s}) \rho_p(r) + 
f^{P}_{\bar{p}n}(\delta \sqrt{s}) \rho_n(r) \right)~.
\end{equation}
Here, $f^P_{\bar{p}p}(\delta \sqrt{s})$ and $f^P_{\bar{p}n}(\delta \sqrt{s})$ denote the $P$-wave 
$\bar{p}p$ and $\bar{p}n$ free-space scattering amplitudes, respectively. We assume that the $P$-wave 
interaction contributes mainly near the nuclear surface where the nuclear densities are relatively low,   
and further in the interior its effect should decrease due to gradient form of the $P$-wave potential. 
Therefore, we do not consider medium modifications of the Paris $P$-wave amplitudes. 
The free-space $\bar{p}p$ and $\bar{p}n$ $P$-wave scattering amplitudes derived 
from the latest version of the Paris $\bar{N}N$ potential are shown as a function 
of energy in Fig.~\ref{fig.:3}.  Again, we witness a strong energy dependence of the amplitudes. 

The analysis of Friedman et al.~\cite{friedmanNPA15} revealed that the potential constructed 
from the Paris $S$- and $P$-wave amplitudes fails to fit the antiproton atom data and that it is 
through the fault of the $P$-wave part. Their analysis also showed that the potential based on the Paris $S$-wave 
and phenomenological $P$-wave amplitude $f^P_{{\bar p}N}=2.9+i1.8$~fm$^3$ \cite{friedmanNPA15} does fit the data well. 
Therefore, we performed calculations exploring the effect of the $P$-wave interaction using both the Paris and 
phenomenological $P$-wave interactions.  

\section{Results}

In this section, we present selected results of our self-consistent calculations of $\bar{p}$ quasi-bound states 
in nuclei across the periodic table using an optical potential constructed from the $\bar{p}N$ scattering amplitudes 
 derived from the 2009 version of the Paris ${\bar N}N$ potential \cite{paris2}. 
First, we performed calculations using only the $S$-wave optical potential and explored its energy and 
density dependence. Then, we took into account the $P$-wave $\bar{p}N$ interaction and studied its effect on the 
$\bar{p}$ binding energies and widths.\\ We performed static, as well as dynamical calculations. 
In the static calculations, the nuclear core is unaffected by the presence of the antiproton and its structure 
thus remains the same. In the dynamical calculations, the $\bar{p}$ polarizes the nuclear core, causing changes 
in the nuclear density distribution and nucleon single-particle energies. In our previous calculations of 
$\bar{p}$ quasi-bound states within the RMF model \cite{hmNPA16} it was demonstrated that the nuclear core is 
significantly affected by the extra antiproton --- the nuclear density in the central region reaches $2-3$ 
times the saturation density. Since the ${\bar p}$ optical potential is density dependent, such increase in 
the density would result in a considerable increase of the $\bar{p}$ binding energies and widths. In fact, 
there is a competing effect, energy dependence of the imaginary part of the phenomenological ${\bar p}N$ scattering amplitude, coming from the phase space suppression     
for the ${\bar p}$ annihilation products, 
which partly compensates the effect of the increased density. The corresponding lifetime of the $\bar{p}$ 
inside a nucleus is then $\sim 1$~fm/c~\cite{hmNPA16}. However, the response of the nuclear core to the extra 
$\bar{p}$ is not instant --- it could possibly last longer than the lifetime of $\bar{p}$ inside a nucleus \cite{larionovPRC08, lmsgPRC10}. 
As a result, the antiproton annihilates before the nuclear core is fully polarized. 
Our static and dynamical calculations of $\bar{p}$ binding 
energies and widths may thus be considered as two limiting scenarios.

\begin{figure}[t!]
\begin{center}
\includegraphics[width=0.47\textwidth]{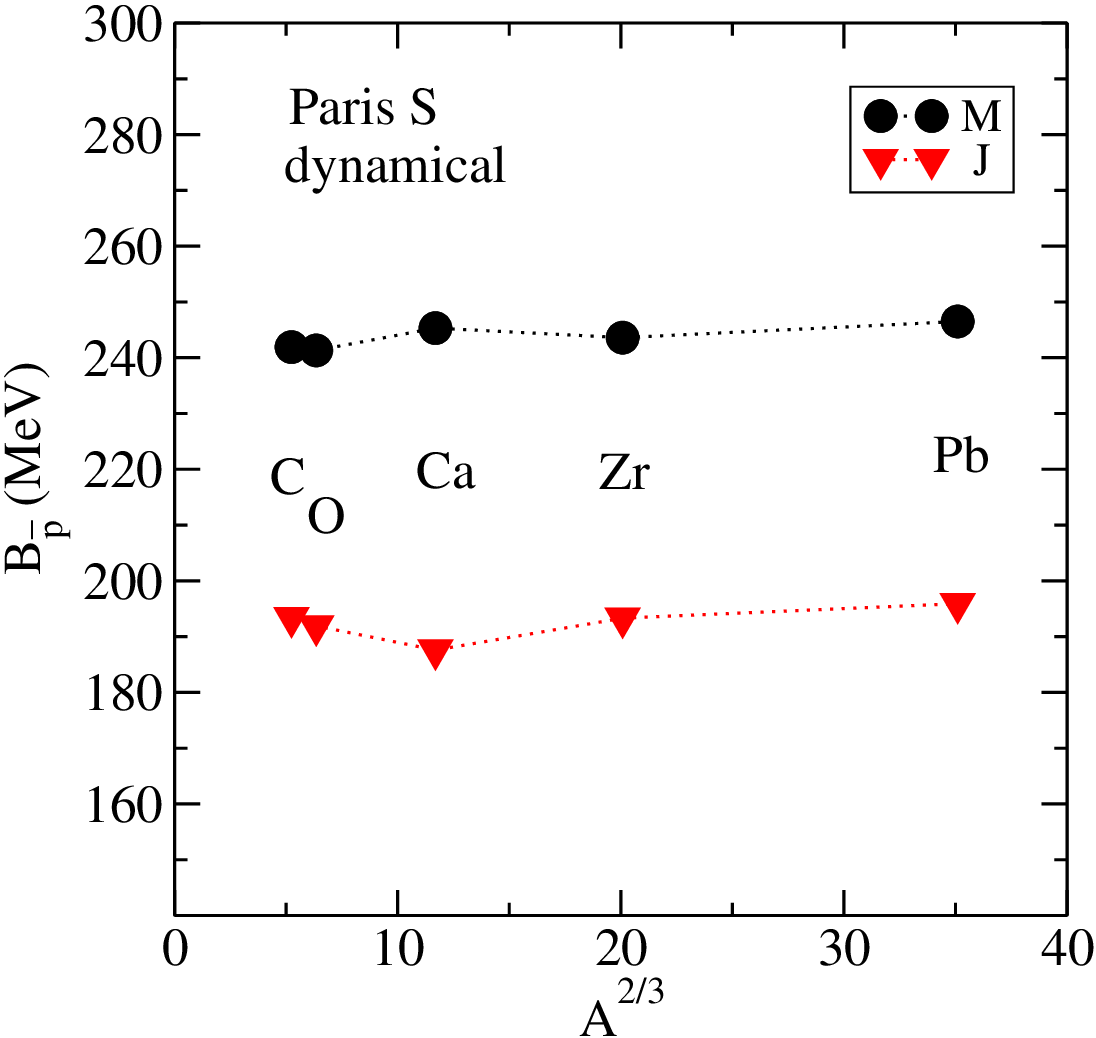} \hspace{10pt}
\includegraphics[width=0.47\textwidth]{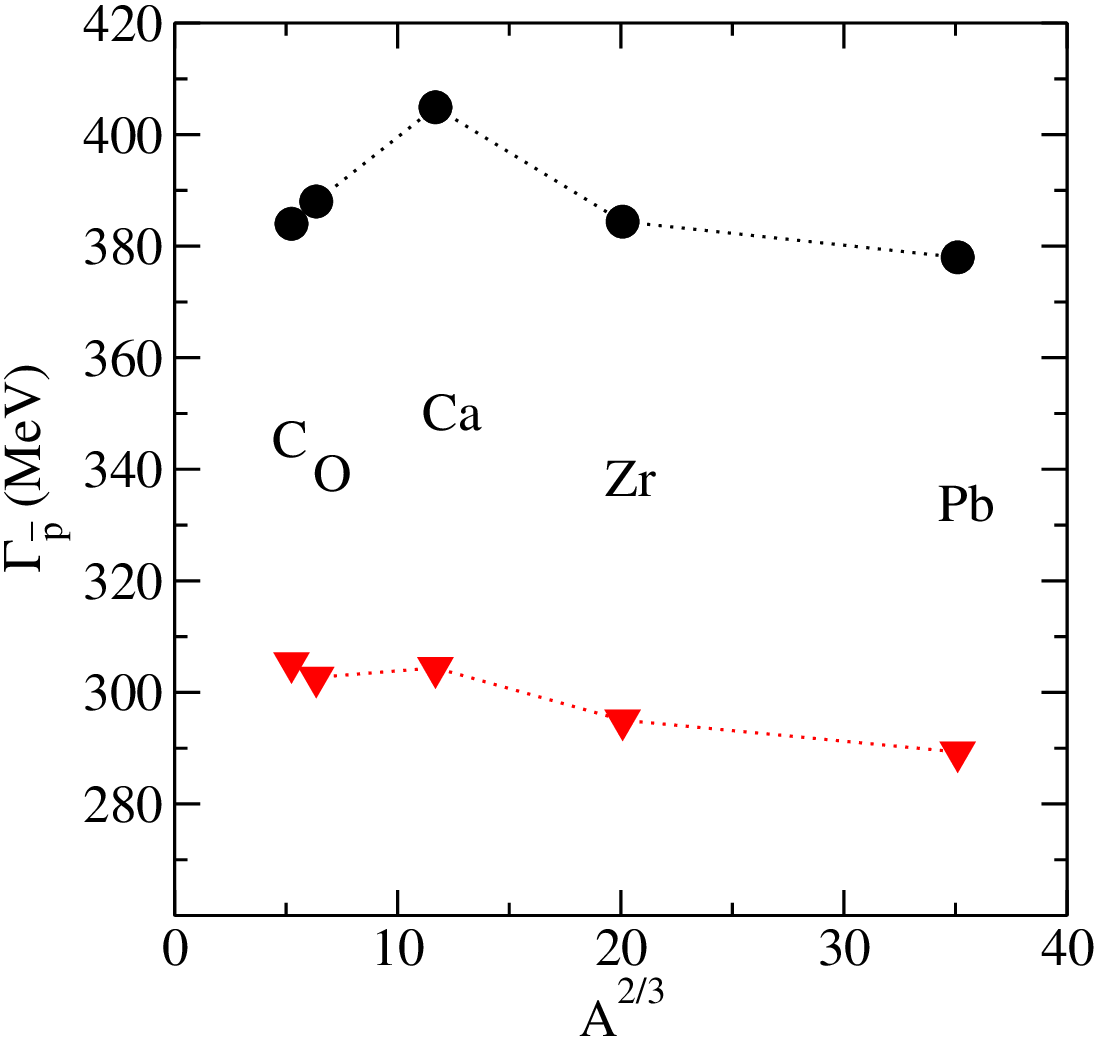}
\caption{\label{fig.:5}Binding energies (left panel) and widths (right panel) of $1s$ $\bar{p}$ bound 
states in selected nuclei, calculated dynamically with $S$-wave Paris potential and different forms 
of $\sqrt{s}$ [M (J) denotes $\sqrt{s}_{\rm M}$ ($\sqrt{s}_{\rm J}$), see text for details].}
\end{center}
\end{figure}

As was shown in Figs.~\ref{fig.:1} and \ref{fig.:3}, the $\bar{p}N$ scattering amplitudes strongly depend 
on energy. It is thus very important to evaluate the $\bar{p}$-nucleus potential self-consistently in the 
appropriate reference frame. This is demonstrated in Fig.~\ref{fig.:5}, where 
we present $1s$ $\bar{p}$ binding energies (left panel) and corresponding widths (right panel) in various 
nuclei calculated dynamically using the Paris $S$-wave potential and two forms of the energy factor:  
$\sqrt{s}_{\rm M}$ [Eq.~\eqref{Eq.:M}] and $\sqrt{s}_{\rm J}$ [Eq.~\eqref{Eq.:J}].
The two forms of $\sqrt{s}$ yield very different binding energies and widths. As for the $\sqrt{s}_{\rm M}$, 
the $\bar{p}$ binding energies are sizable and show weak $A$-dependence. The corresponding $\bar{p}$ widths are 
huge ($\leq 400$~MeV), much larger than the binding energies. 
Including momentum dependent terms in $\sqrt{s}_{\rm J}$ causes additional considerable downward energy shift,   
which leads to  lower values of $\bar{p}N$ scattering amplitudes (see Fig.~\ref{fig.:1}; the relevant 
energy shift $\delta \sqrt{s}\leq -200$~MeV) and, consequently, 
shallower $\bar{p}$-nucleus optical potential. The ${\bar p}$ widths are strongly reduced, yet remain sizable. 
The $\bar{p}$ binding energies decrease as well (up to $\sim 20$\%) and are again almost $A$-independent. 
Finally, it is to be noted that in static calculations, 
the effect of the momentum dependent terms in $\sqrt{s}_{\rm J}$ 
on $\bar{p}$ binding energies and widths is about half of that effect in the dynamical case.  

Next, we consider the $P$-wave part of the $\bar{p}N$ interaction. 
 We adopt the Paris $\bar{p}p$ and $\bar{p}n$ $P$-wave scattering amplitudes as well as the phenomenological 
$P$-wave potential fitted by Friedman and Gal to $\bar{p}$ atom data \cite{friedmanNPA15} and 
construct the $S+P$-wave $\bar{p}$-nucleus optical potential [Eq.~\eqref{SPpot}] which is further 
applied in self-consistent calculations of ${\bar p}$-nuclear quasi-bound states.

\begin{figure}[t!]
\begin{center}
\includegraphics[width=0.47\textwidth]{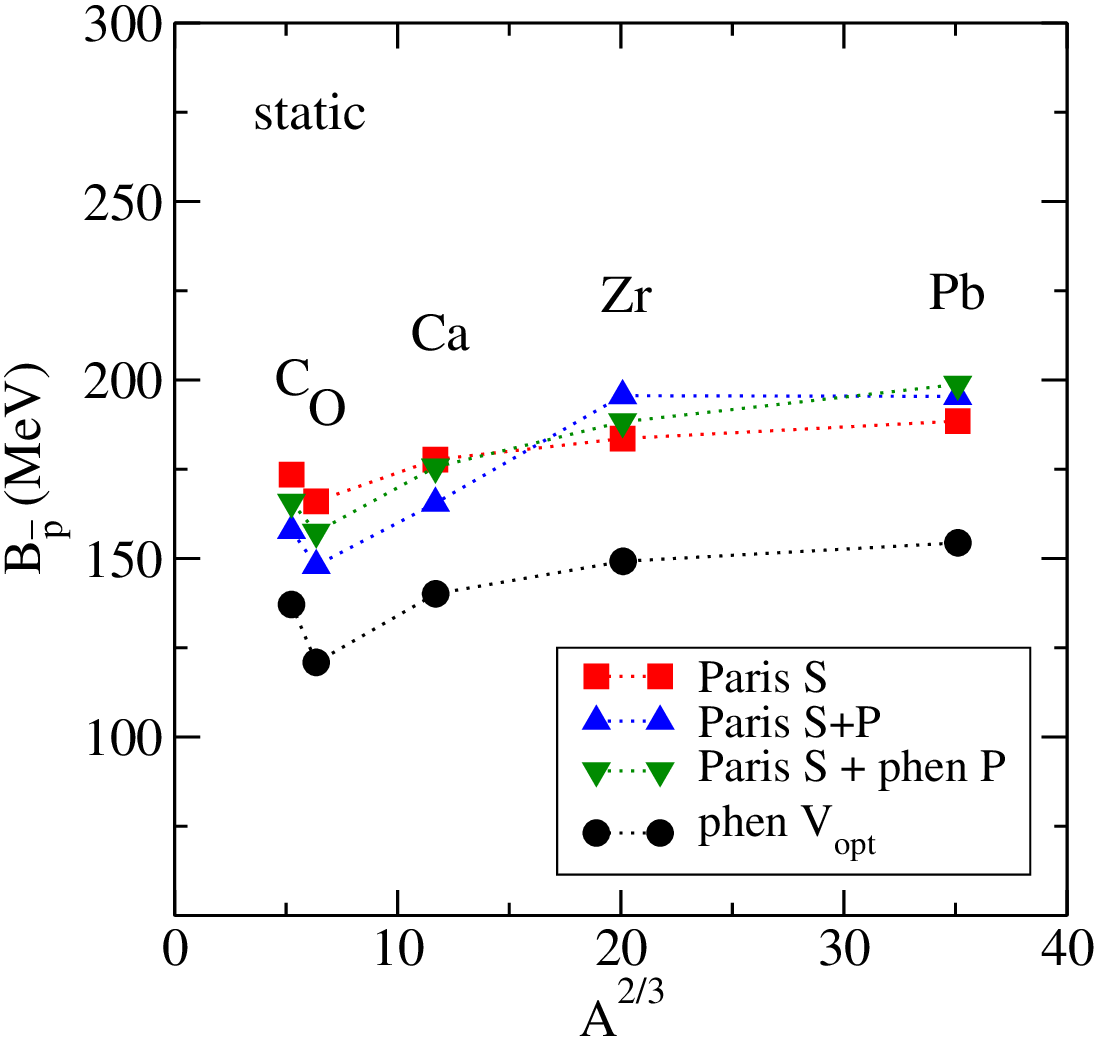} \hspace{10pt}
\includegraphics[width=0.47\textwidth]{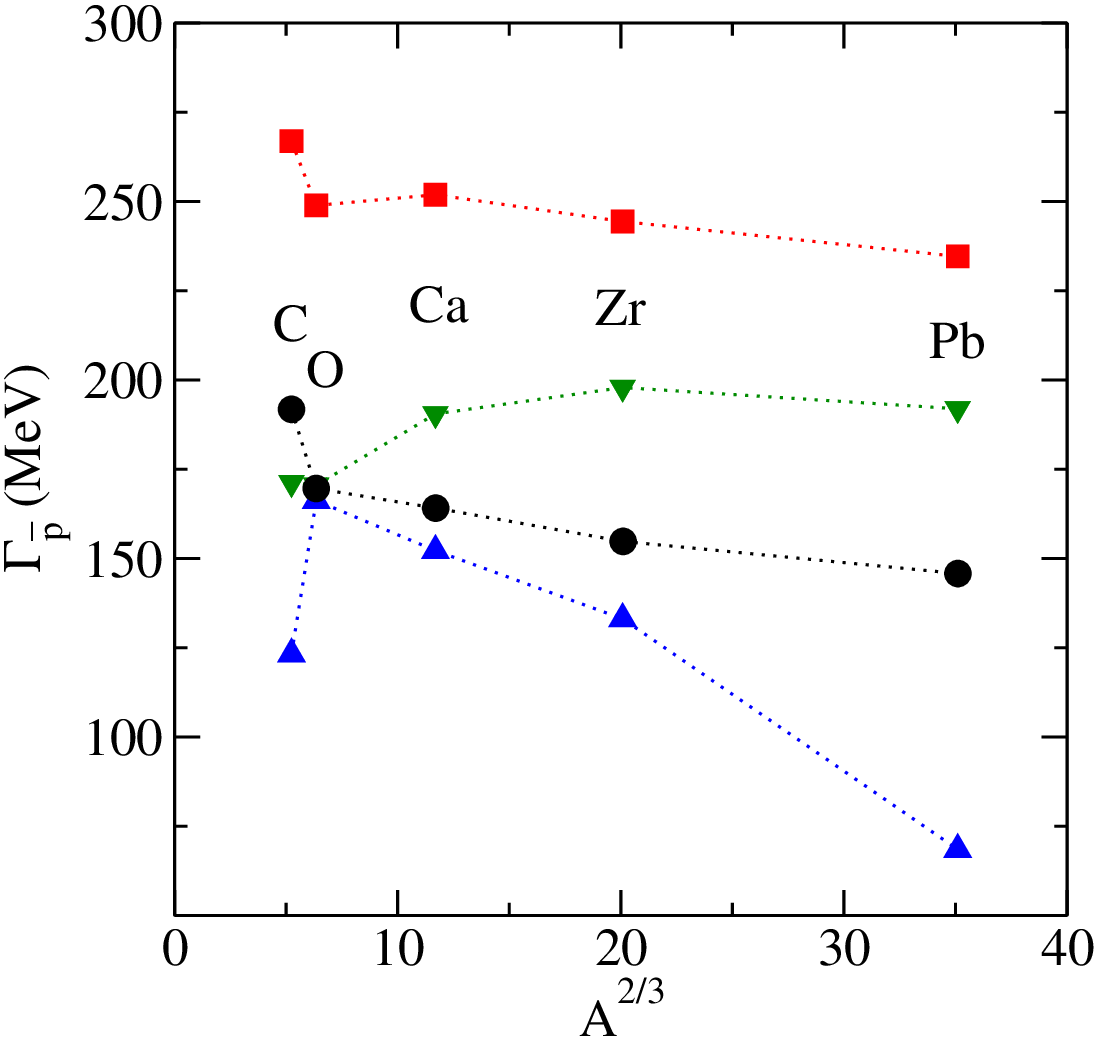}
\caption{\label{fig.:6}
$1s$ $\bar{p}$ binding energies (left panel) and widths (right panel) in various nuclei, calculated
statically for $\sqrt{s}_{\rm J}$ using $S$-wave Paris potential (red squares), including phenomenological 
$P$-wave potential (green triangles down), Paris $P$-wave potential (blue triangles up) and phenomenological 
RMF potential (black circles).}
\end{center}
\end{figure}

\begin{figure}[t!]
\begin{center}
\includegraphics[width=0.65\textwidth]{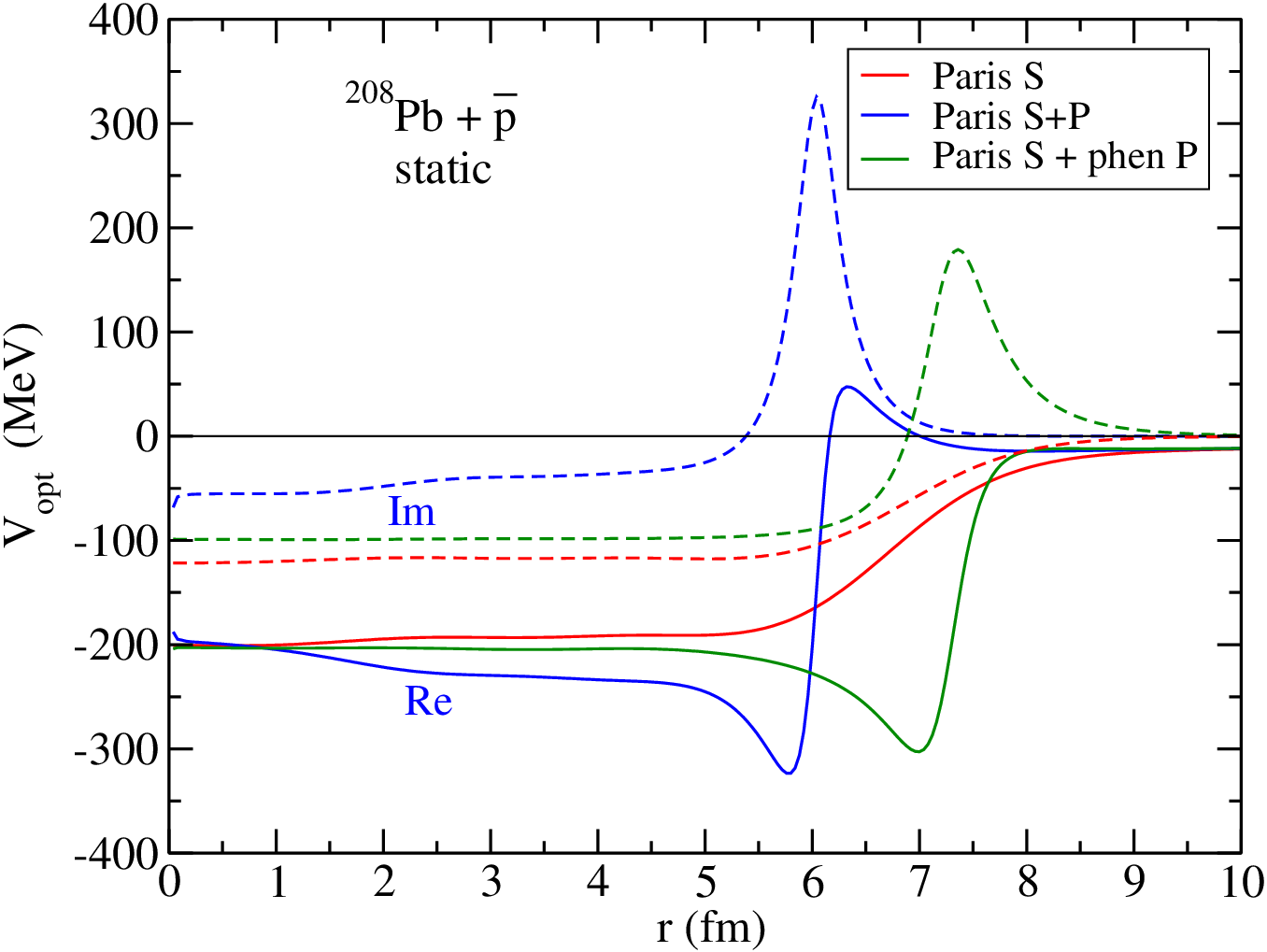} 
\caption{\label{fig.:11}
The real (solid curves) and imaginary (dashed curves) parts of the $S$-wave Paris potential (red) and the local (Krell-Ericson \cite{keNPB69}) forms of the Paris 
$S+P$-wave (green) and Paris $S$-wave + phen. $P$-wave (blue) potentials felt by $\bar{p}$ in  
$^{208}$Pb, calculated statically for $\sqrt{s}_{\rm J}$ (see text for details).}
\end{center}
\end{figure}

In Fig.~\ref{fig.:6}, we present $1s$ $\bar{p}$ binding energies (left) and widths (right) as a function of 
mass number $A$, calculated statically with the Paris $S$-wave (squares), Paris $S+P$-wave (triangles up), 
Paris $S$-wave + phen. $P$-wave (triangles down) potentials for $\sqrt{s}_{\rm J}$~[Eq.~\eqref{Eq.:J}]. 
The $\bar{p}$ binding energies and widths calculated statically with a 
phenomenological optical potential (`phen $V_{\rm opt}$', circles)~\cite{hmNPA16} are shown for comparison.  
The real part of this $\bar{p}$-nucleus potential was constructed within the RMF approach using 
G-parity motivated $\bar{p}$--meson coupling constants which were multiplied by a scaling factor to account 
for available experimental data. The $\bar{p}$ absorption was described by the imaginary part of a purely 
phenomenological optical potential fitted to strong interaction energy shifts and widths in ${\bar p}$-atoms. 
The reduced phase space available for annihilation of ${\bar p}$ deeply bound in the nuclear medium  
was taken into account by introducing corresponding suppression factors (see Ref.~\cite{hmNPA16} for more details). 
\\ 
As can be seen from Fig.~\ref{fig.:6}, 
both $P$-wave interaction terms, Paris as well as phenomenological, do not affect much the $\bar{p}$ binding 
energies --- they are comparable with binding energies evaluated using only the $S$-wave potential. 
On the other hand, the $\bar{p}$ widths decrease noticeably when the phenomenological $P$-wave term is 
included in the ${\bar p}$ optical potential. 
The effect is even more pronounced for the Paris $P$-wave interaction. We observe strong $A$-dependence 
of $\bar{p}$ widths for the Paris $S+P$-wave potential. On the contrary, the widths calculated with the 
phenomenological $P$-wave term, as well as only with the $S$-wave potential vary much less with $A$ 
(starting oxygen).\\ 
\begin{table}[t!]
\caption{Self-consistent energy shifts $\delta \sqrt{s}_{\rm J}$ in $^{208}$Pb+$\bar{p}$ relevant to static calculations 
within the Paris $S$-wave, Paris $S+P$-wave and Paris $S$-wave + phen. $P$-wave potentials.}
\vspace*{10pt}
\begin{center}
 \begin{tabular}{l|ccc}
\hline
  $^{208}$Pb+$\bar{p}$ & Paris $S$ & Paris $S+P$ & Paris $S$ + phen. $P$ \\ \hline
  $\delta \sqrt{s}_{\rm J}$ (MeV) & -210.6 & -238.9 & -223.6\\ 
 \hline
 \end{tabular}
 \end{center}
 \label{Tab.:3}
\end{table}  
To better understand this behavior, we show in Table~\ref{Tab.:3} the energy shifts 
$\delta \sqrt{s}_{\rm J}$ in $^{208}$Pb+$\bar{p}$ evaluated self-consistently in static calculations 
with Paris $S$-wave, Paris $S+P$-wave and Paris $S$-wave + phen. $P$-wave potentials.  
The $S$-wave potential yields the smallest energy shift with 
respect to threshold, which implies stronger $\bar{p}N$ amplitudes (see Fig.~\ref{fig.:1}) and thus 
larger $\bar{p}$ binding energies and widths. When the $P$-wave interaction is taken into account, the 
downward energy shift increases. As a result, the $S$-wave part of the ${\bar p}$ potential becomes weaker. 
However, this decrease of the $S$-wave attraction is more than compensated by the real part of the $P$-wave 
potential as illustrated in Fig.~\ref{fig.:11}.
 Here we present the Paris $S$-wave,  Paris 
$S+P$-wave and Paris $S$-wave + phen. $P$-wave ${\bar p}$ potentials in $^{208}$Pb, calculated statically for 
$\sqrt{s}_{\rm J}$~\footnote{Fig.~\ref{fig.:11} shows local forms of the $S+P$-wave potentials obtained from nonlocal 
Kisslinger potential of Eq.~\eqref{SPpot} using the Krell-Ericson transformation~\cite{keNPB69}.}.  
As a result, the ${\bar p}$ binding energies shown in Fig.~\ref{fig.:6} are very close to each other. 
On the other hand, the weaker imaginary part of the $S$-wave potential is not fully compensated by the $P$-wave part, 
particularly the Paris $P$-wave which is very weakly absorptive for the corresponding $\delta \sqrt{s}_{\rm J}$ 
(see Fig.~\ref{fig.:3} and Table~\ref{Tab.:3}). The imaginary part of the $S+P$ potential is  
thus shallower than that of pure $S$-wave potential. 
On top of that, the range of the Paris $S+P$-wave potential is smaller than the range 
of the Paris $S$-wave + phen. $P$-wave potential (see Fig.~\ref{fig.:11}). Therefore, the $\bar{p}$ widths in 
heavier nuclei calculated using the Paris $S+P$-wave potential decrease considerably.\\ 
It is to be noted that the depth of the $S+P$-wave potential is a result of delicate interplay between the 
$S$- and $P$-wave parts which are linked together. Very important is also the balance between the real and imaginary 
parts of the $P$-wave amplitudes since their strength controls the range of the potential. 

\begin{figure}[t!]
\begin{center}
\includegraphics[width=0.47\textwidth]{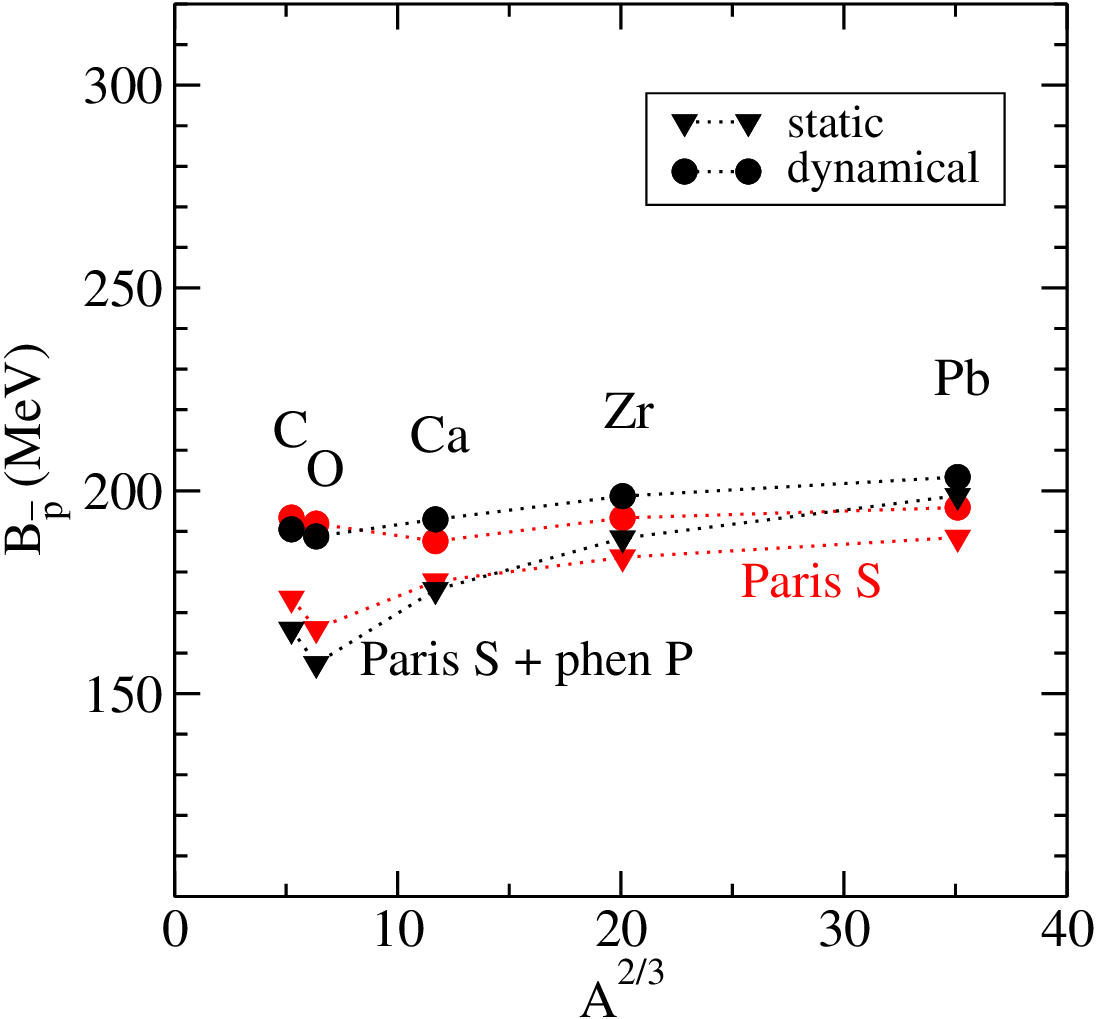} \hspace{10pt}
\includegraphics[width=0.47\textwidth]{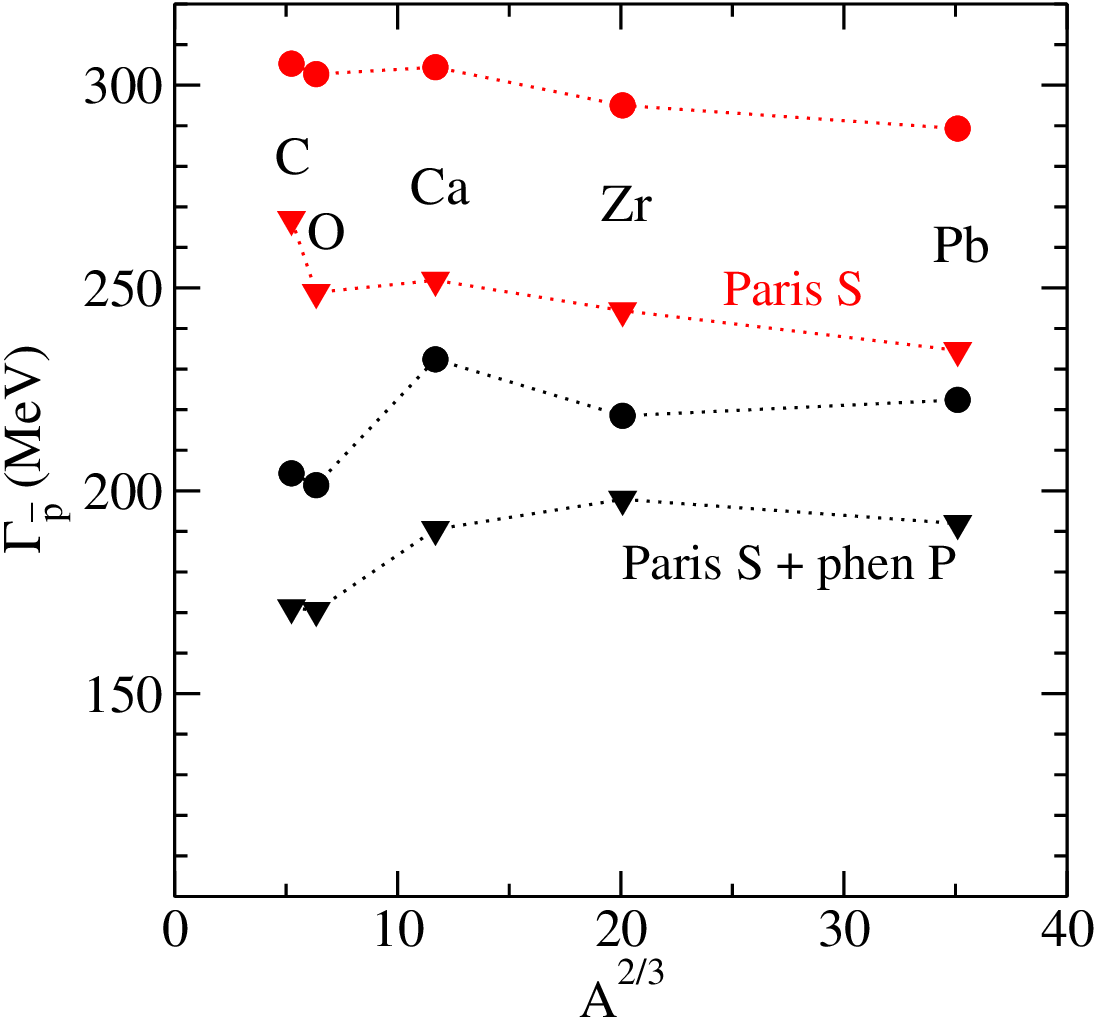}
\caption{\label{fig.:7}
$1s$ $\bar{p}$ binding energies (left panel) and widths (right panel) in various nuclei, calculated
statically (triangles) and dynamically (circles) for $\sqrt{s}_{\rm J}$ using $S$-wave Paris potential (red)  
and including phenomenological $P$-wave potential (black).}
\end{center}
\end{figure}

Dynamical effects are illustrated in Fig.~\ref{fig.:7} where we compare $1s$ $\bar{p}$ binding energies 
(left panel) and corresponding widths (right panel) in various nuclei, calculated statically  
and dynamically for $\sqrt{s}_{\rm J}$ using the Paris $S$-wave and Paris $S$-wave + phen. $P$-wave optical 
potentials.   
In both cases, the binding energies $B_{\bar p}$ calculated dynamically are somewhat larger than those 
obtained in static calculations and the polarization effects decrease with the mass number $A$. 
In dynamical and static calculations alike, the $\bar{p}$ binding energies calculated for the 
Paris $S$ + phen. $P$-wave potential are comparable with those obtained with the Paris $S$-wave potential.   
The $P$-wave interaction slightly increases the $\bar{p}$ binding energies in 
heavier nuclei ($^{40}$Ca, $^{90}$Zr, and $^{208}$Pb) and decreases them in light nuclei 
($^{16}$O and $^{12}$C). 
The $\bar{p}$ widths calculated dynamically are noticeably larger than the widths calculated statically. 
It is caused mainly by the increase of the central nuclear density, which outweighs the decrease of the 
${\bar p}N$ amplitudes due to the larger energy shift with respect to threshold 
($\delta \sqrt{s}_{\rm dyn} \sim -255$~MeV vs. $\delta \sqrt{s}_{\rm stat} \sim -200$~MeV).   
The Paris $S$-wave + phen. $P$-wave potential yields again smaller $\bar{p}$ 
widths than the $S$-wave potential. Still, the $\bar{p}$ widths calculated dynamically are larger or at least  
comparable with the corresponding $\bar{p}$ binding energies. The lifetime of the antiproton inside the 
nucleus is consistent with $\simeq 1$~fm/c.\\
It is to be noted that in our previous RMF calculations \cite{hmNPA16} we found strong model dependence of the dynamical effects caused by the extra $\bar{p}$ inside the nucleus. It could be attributed to different values of nuclear compressibility given by applied RMF models (models with larger compressibility predict larger dynamical changes in $\bar{p}$ binding energies). In order to explore model dependence in the present study, we performed calculations also using the RMF model TM(1)2~\cite{Toki}. 
We found that unlike the phenomenological RMF approach the present static as well as dynamical calculations based on Paris $\bar{N}N$ amplitudes yield quite similar results within the TM and NL-SH models, the differences in $\bar{p}$ binding energies 
and widths are up to 10~MeV. It is due to energy dependence of the $\bar{p}N$ amplitudes which compensates the increase of the nuclear density. Namely, larger dynamical changes imply larger subthreshold energy shift and thus weaker $\bar{p}N$ amplitudes (see Fig.~\ref{fig.:1}). We preferred the NL-SH model in the present work since the TM model consists of two 
different parameter sets -- TM2 for light nuclei and TM1 for heavy nuclei. 
\begin{figure}[b!]
\begin{center}
\includegraphics[width=0.47\textwidth]{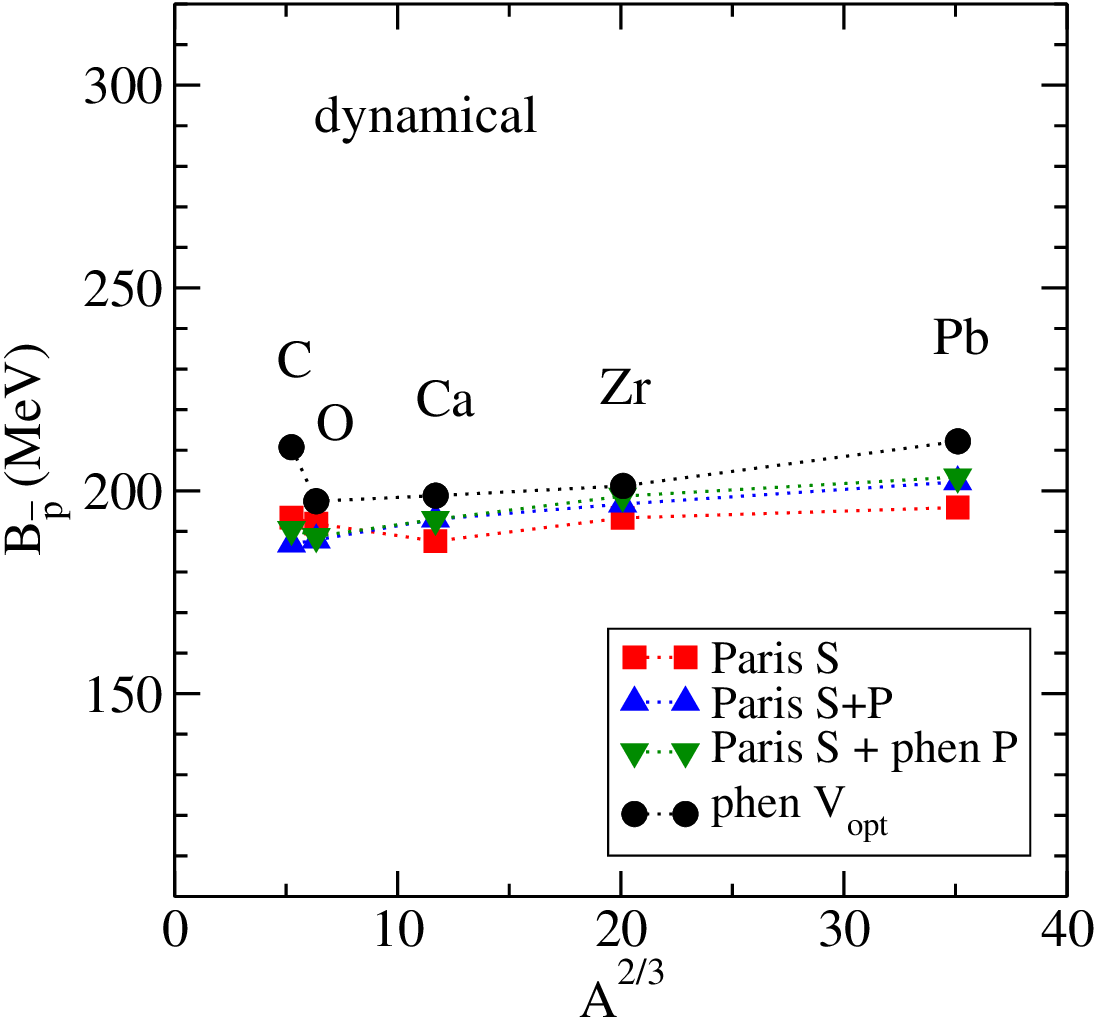} \hspace{10pt}
\includegraphics[width=0.47\textwidth]{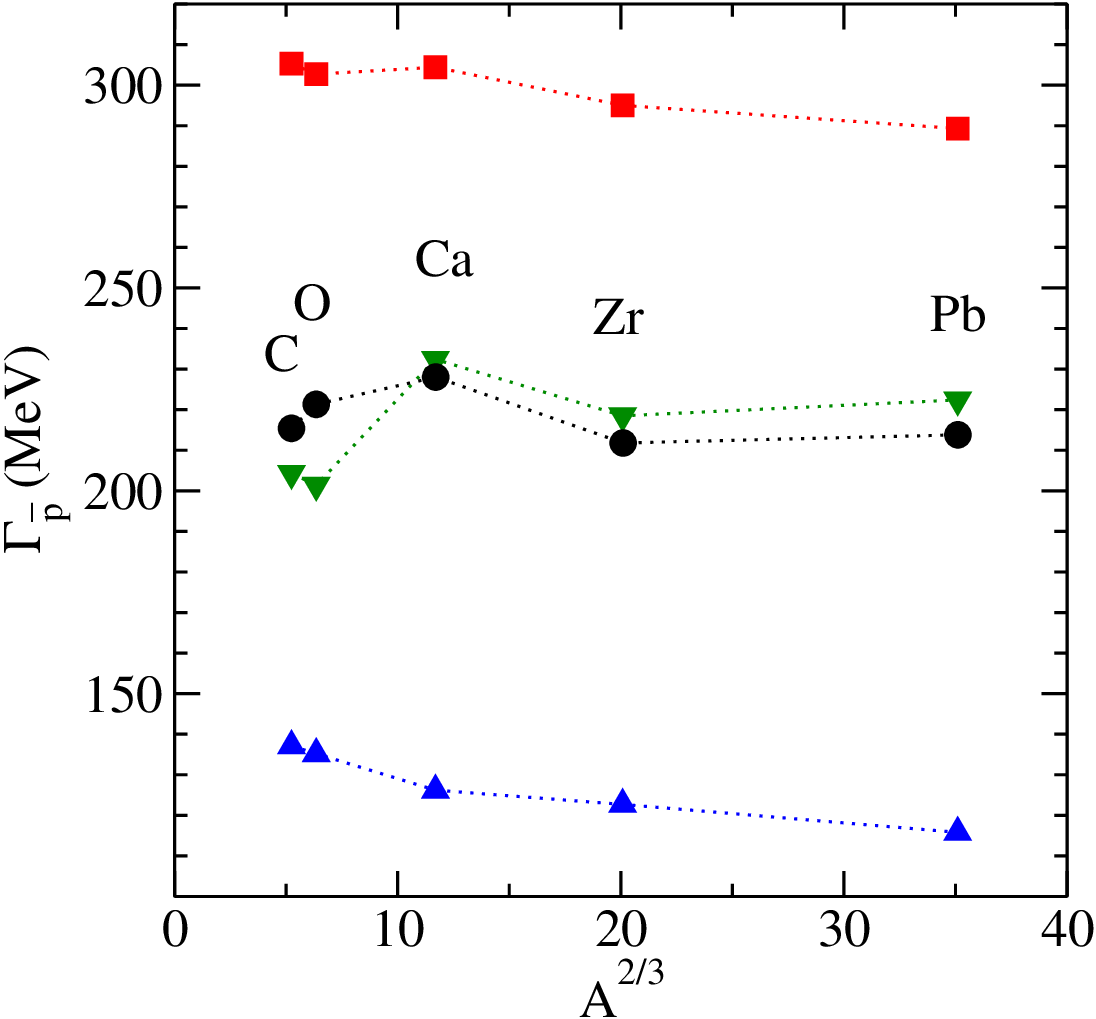}
\caption{\label{fig.:8}
Binding energies (left panel) and widths (right panel) of $1s$ $\bar{p}$-nuclear states in selected nuclei,
calculated dynamically for $\sqrt{s}_{\rm J}$ using the Paris $\bar{N}N$ $S$-wave 
potential (red), Paris $S$-wave + phen. $P$-wave (green) and phenomenological approach 
within the RMF model NL-SH (black).}
\end{center}
\end{figure}

Next, we compare the predictions for $\bar{p}$ binding energies and widths calculated dynamically using the 
2009 version of the  Paris $\bar{N}N$ potential with our former calculations based on the RMF model \cite{hmNPA16}. 
The $1s$ $\bar{p}$ binding energies (left) and corresponding widths (right) in selected nuclei calculated 
using the phenomenological RMF approach (circle), Paris $S$-wave potential (square), Paris $S+P$-wave potential 
(triangle up) and Paris $S$-wave + phen. $P$-wave potential (triangle down) are shown in Fig.~\ref{fig.:8}. 
The binding energies are very close to each other in all cases, $B_{\bar p}\sim 200$~MeV, and rather 
weakly $A$-dependent. The $\bar{p}$ widths exhibit considerably larger dispersion for the different potentials. 
The Paris $S$-wave potential yields sizable widths in all nuclei, $\Gamma_{\bar p}\sim 300$~MeV. 
The Paris $P$-wave interaction again reduces the $\bar{p}$ widths significantly, to less than one half. 
The Paris $S$-wave + phen. $P$-wave potential yields very similar $\bar{p}$ widths as the phenomenological 
approach. They are in the range of $\sim 200 - 230$~MeV and comparable with the corresponding binding energies. 
The agreement between the phenomenological RMF and Paris $S$-wave + phen. $P$-wave potentials is quite impressive.\\  
One has to mention that in the dynamical calculations, the depths of the Paris $S+P$-wave and Paris $S$-wave + 
phen. $P$-wave potentials in the central region of all nuclei are very similar to each other. However, the 
range of the Paris $S+P$-wave potential (in the local form) is again much smaller than the range of the Paris $S$ + 
phen. $P$-wave potential. Consequently, the $\bar{p}$ widths calculated using the Paris $S+P$ potential are considerably 
smaller.\\  
We may thus infer that the real 
and imaginary parts of the Paris $P$-wave amplitudes are not well balanced in the energy region relevant to 
${\bar p}$-nuclear states calculations. 
Anyway, it was demonstrated by Friedman and Gal (see Table~1 in Ref.~\cite{friedmanNPA15}) that the real and 
imaginary parts of the Paris $P$-wave had to be scaled by different factors in order to obtain satisfactory 
fit to $\bar{p}$ atom data. 

\begin{figure}[t!]
\begin{center}
\includegraphics[width=0.94\textwidth]{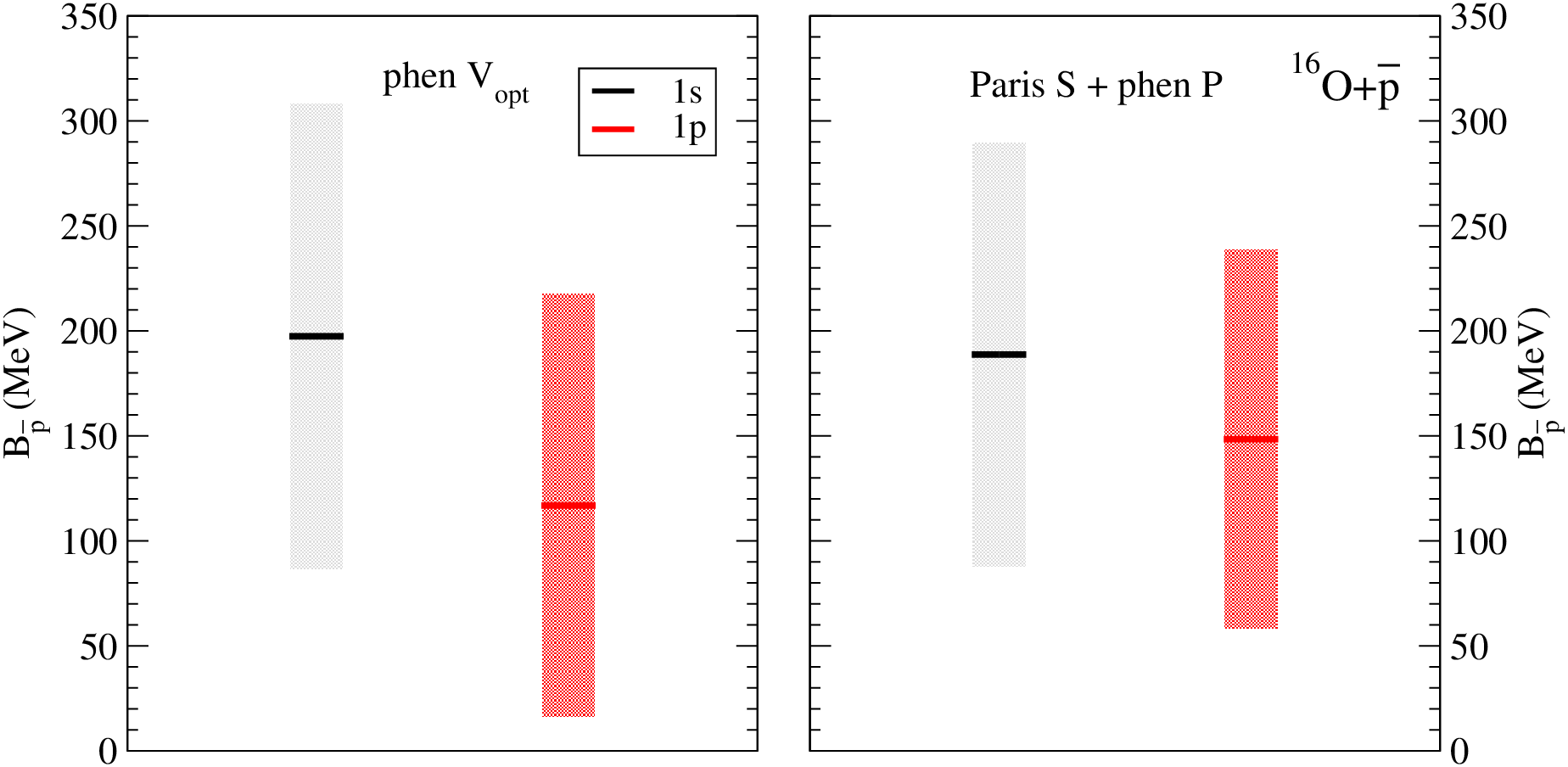}
\caption{\label{fig.:9}
$1s$ and $1p$ binding energies (lines) and widths (boxes) of $\bar{p}$ in $^{16}$O calculated dynamically 
within the NL-SH model for $\sqrt{s}_{\rm J}$ with phenomenological $\bar{p}$ optical potential (left panel) 
and Paris $S$-wave + phen. $P$-wave potential (right panel).}
\end{center}
\end{figure}
Besides the $\bar{p}$ ground states we calculated also $\bar{p}$ excited states in selected nuclei. 
In Fig.~\ref{fig.:9} we compare the binding energies and widths of the $1s$ and $1p$ $\bar{p}$ states 
in $^{16}$O, calculated dynamically using the Paris $S$-wave + phen. $P$-wave potential (right) and within the 
phenomenological RMF approach (left). The Paris $S$-wave + phen. $P$-wave potential yields similar spectrum of 
the $\bar{p}$ bound states as the RMF potential, however the $1p$ binding energy is 
about $20\%$ larger and the width is slightly smaller than in the RMF model. 
Nevertheless, the agreement of the two spectra, which were obtained within two different approaches, is 
surprisingly good. 

\begin{figure}[t!]
\begin{center}
\includegraphics[width=0.94\textwidth]{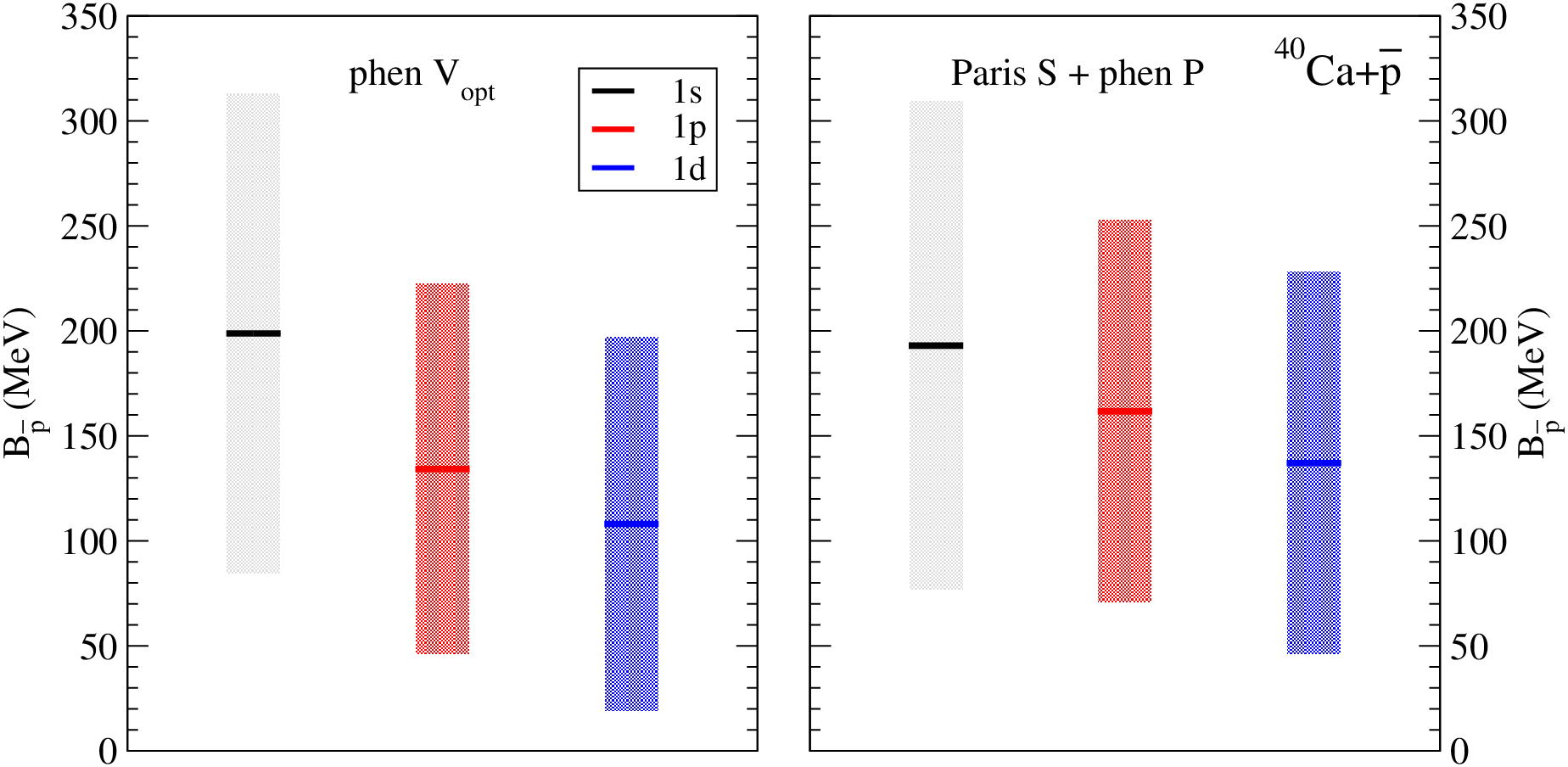}
\caption{\label{fig.:10}
$1s$, $1p$ and $1d$ binding energies (lines) and widths (boxes) of $\bar{p}$ in $^{40}$Ca calculated 
dynamically within the NL-SH model for $\sqrt{s}_{\rm J}$ with phenomenological $\bar{p}$ optical potential 
(left panel) and Paris $S$-wave + phen. $P$-wave potential (right panel).}
\end{center}
\end{figure}
Fig.~\ref{fig.:10} shows similar ${\bar p}$ spectra in $^{40}$Ca. The $1p$ and $1d$ binding energies 
calculated with the Paris $S$-wave + phen. $P$-wave potential are again slightly larger 
(and $s-p$ and $s-d$ level spacing smaller) than 
in the phenomenological RMF approach. It is due to a broader $\bar{p}$ potential well 
of the Paris $S$-wave + phen. $P$-wave potential. Both approaches yield comparable 
$\bar{p}$ widths. 
\\
It is to be noted that in the present calculations, $V_{\rm opt}$ is a central potential constructed from angular momentum-averaged scattering amplitudes and thus there is no spin-orbit splitting of $p$ and $d$ levels presented in Figs.~\ref{fig.:9} and \ref{fig.:10}. 
In the RMF approach, the $\bar{p}$ binding energies 
in $1p$ and $1d$ spin doublets are nearly degenerate, the difference in $\bar{p}$ energies (as well as $\bar{p}$ widths) is up to~$\sim 1$~MeV. These findings are in agreement with spin symmetry predicted for 
antinucleon spectra \cite{ginocchio, heEPJ06, llmgEPJ10, lmafcPRC10}. In the left panels of Figs.~\ref{fig.:9} and \ref{fig.:10} we show  spin-averaged $1p$ and $1d$ $\bar{p}$ binding energies and widths.

\section{Summary}
We performed fully self-consistent calculations of $\bar{p}$-nuclear quasi-bound states using an optical 
potential constructed from the $S$- and $P$-wave ${\bar p}N$ scattering amplitudes obtained within 
the 2009 version of the Paris $\bar{N}N$ potential~\cite{friedmanNPA15}. 
The free-space $S$-wave scattering amplitudes were modified by WRW procedure~\cite{wrw} in order to account for
 Pauli correlations in the medium. A self-consistent scheme for proper dealing with the energy and density 
dependence of the in-medium amplitudes was adopted in evaluation of the $\bar{p}$-nuclear optical potential.
To our knowledge, such calculations based on a microscopic model were carried out for the first time. 
Previous studies of $\bar p$-nuclear states were performed within phenomenological (RMF) approaches \cite{burvenich,Mishustin,larionovPRC08,lmsgPRC10,lenske,hmNPA16}. 
     
First, we explored the $S$-wave part of the ${\bar p}$ optical potential and showed that its form  
depends strongly on energy and density at which it is evaluated. The potential derived from free-space $\bar{p}N$ amplitudes is repulsive and moderately absorptive 
at threshold. After applying in-medium modifications of the amplitudes, the potential becomes strongly attractive and absorptive at energies and densities relevant to 
${\bar p}$-nuclear states calculations. As a result, ${\bar p}$ binding energies in the $1s$ state amount to  
almost $200$~MeV, and the corresponding widths $\Gamma_{\bar p} \sim 300$~MeV in the dynamical calculations.   

Then we took into 
account the $P$-wave part of the ${\bar p}$ optical potential. Recent analysis by 
Friedman et al.~\cite{friedmanNPA15} revealed that the optical potential based on the  Paris $S$- 
and $P$-wave scattering amplitudes fails to fit the $\bar{p}$ atom data. On the other hand, the 
Paris $S$-wave potential supplemented by a phenomenological $P$-wave term reproduces the data well. 
We adopted both the Paris and phenomenological $P$-wave terms in our calculations. 
We performed static calculations (neglecting modifications of the nuclear core) as well as dynamical  
calculations (nuclear core is polarized by $\bar{p}$) which yield lower and upper estimates of 
$\bar{p}$ binding energies and widths. We found that the $P$-wave interaction almost does not 
affect the binding energies of ${\bar p}$-nuclear quasi-bound states. 
This is in sharp contrast to the case of ${\bar p}$ atoms where it was found necessary to include the $P$-wave term 
of the Paris ${\bar p}N$ interaction in  order to increase attraction of the ${\bar p}$ optical 
potential~\cite{friedmanNPA15}. This again illustrates how the form of the potential depends on energy and 
density.\\   
The widths of ${\bar p}$-nuclear states are reduced substantially when the $P$-wave 
interaction part is considered. The Paris 
$P$-wave potential reduces the widths much more than the phenomenological one. 
It is a result of a delicate balance between the $S$- and $P$-wave parts of the total ${\bar p}$ optical 
potential. The strength of the $P$-wave part which acts mainly near the nuclear surface and thus 
controls the range of the optical potential seems to be decisive. 

Finally, we compared results of our present calculations using the Paris ${\bar N}N$ potential with our previous 
calculations of ${\bar p}$-nuclear quasi-bound states performed within the RMF model tuned to the ${\bar p}$-atom data ~\cite{hmNPA16}. 
The ${\bar p}$ binding energies and widths calculated dynamically with the Paris $S$-wave potential 
supplemented by the 
phenomenological $P$-wave term were found to be in good agreement with the RMF model calculations. 
Both approaches yield the $1s$ ${\bar p}$ binding energies $B_{\bar p}\approx 200$~MeV and the widths 
$\Gamma_{\bar p} \sim 200 -230$~MeV in considered nuclei.    
We find this agreement rewarding as it shows that the ${\bar p}$ atoms fits 
not only define the form of the ${\bar p}$ optical potential near threshold and at low density region 
but, moreover, quite sufficiently constrain extrapolations to higher 
densities and farther down below threshold --- to the region relevant to ${\bar p}$-nuclear states.  
 
In conclusion, it is to be noted that the present work based on the 2009 version of the Paris 
${\bar N}N$ potential was inspired by the recent study of Friedman and Gal~\cite{friedmanNPA15}. They 
examined this very potential in the analysis of experimental results for 
antiprotonic atoms across the periodic table as well as antinucleon interactions with nuclei 
up to 400~MeV/c. Other realistic ${\bar N}N$ models, such as the Bonn-J\"ulich chiral NNLO~\cite{KHM14} 
and N$^3$LO~\cite{H17} EFT potential models or Zhou-Timmermans model~\cite{ZT}, could be applied in the study of 
${\bar p}$ interactions with the nuclear medium. It would be desirable to perform such calculations and compare 
between different ${\bar N}N$ interaction models.  

\section*{Acknowledgements}
We thank E. Friedman, A. Gal and S. Wycech for valuable discussions, and B. Loiseau for providing us 
with the free $\bar{N}N$ amplitudes. This work was supported by the GACR Grant No.P203/15/04301S.

\end{document}